%% file: ms.tex
\documentclass[twocolumn]{aastex631}

\definecolor{fuchsia}{rgb}{0.54, 0.17, 0.89}
\definecolor{azure}{rgb}{0.0, 0.5, 1.0}
\definecolor{pgreen}{rgb}{0.12, 0.3, 0.17}
\definecolor{alizarin}{rgb}{0.82, 0.1, 0.26}

\newcommand{\kms}{{\rm km~s^{-1}}}
\newcommand{\oi}{[\textrm{O}~\textsc{i}]}

\newcommand{\oii}{[\textrm{O}~\textsc{ii}]}

\newcommand{\oiii}{[\textrm{O}~\textsc{iii}]}
\newcommand{\oiiif}{$[\textrm{O}~\textsc{iii}]\,4363$}

\newcommand{\nii}{[\textrm{N}~\textsc{ii}]}

\newcommand{\hi}{\textrm{H}~\textsc{i}}
\newcommand{\hii}{\textrm{H}~\textsc{ii}}

\newcommand{\sii}{[\textrm{S}~\textsc{ii}]}
\newcommand{\siii}{[\textrm{S}~\textsc{iii}]}
\newcommand{\siiin}{[\textrm{S}~\textsc{iii}]\,{6716+6731}}

\newcommand{\feii}{[\textrm{Fe}~\textsc{ii}]}
\newcommand{\feiii}{[\textrm{Fe}~\textsc{iii}]}
\newcommand{\hei}{\textrm{He}~\textsc{i}}
\newcommand{\heii}{\textrm{He}~\textsc{ii}}

\newcommand{\ariii}{[\textrm{Ar}~\textsc{iii}]}
\newcommand{\ariv}{[\textrm{Ar}~\textsc{iv}]}
\newcommand{\silii}{[\textrm{Si}~\textsc{ii}]}

\newcommand{\simgt}{\,\rlap{\lower 3.5 pt \hbox{$\mathchar \sim$}} \raise
1pt \hbox {$>$}\,}
\newcommand{\simlt}{\,\rlap{\lower 3.5 pt \hbox{$\mathchar \sim$}} \raise
1pt \hbox {$<$}\,}
\newcommand{\Msun}{M_{\odot}}

\newcommand{\logm}{\log M_*/\Msun}

\newcommand{\logoh}{12 + \log {\rm (O/H)}}

\newcommand{\civf}{\textrm{C}~\textsc{iv}}
\newcommand{\ciiif}{\textrm{C}~\textsc{iii}}
\newcommand{\niiif}{\textrm{N}~\textsc{iii}}

\newcommand{\ha}{${\rm H\alpha}$}
\newcommand{\hb}{${\rm H\beta}$}

\newcommand{\pb}{${\rm Pa\beta}$}
\newcommand{\pg}{${\rm Pa\gamma}$}
\newcommand{\pd}{${\rm Pa\delta}$}

\newcommand{\Z}{{\rm [Z/H]}}

\newcommand{\id}{MACSJ1149-WR1} 
\newcommand{\nline}{31} %

\usepackage{amsmath}

\accepted{November 1, 2024}

\submitjournal{ApJ}

\shorttitle{\id\ at $z=2.76$}
\shortauthors{Morishita et al.}

\graphicspath{{./}{figures/}}

\begin{document}

\title{
Dissecting the Interstellar Media of A Wolf-Rayet Galaxy at $z=2.76$
}

\correspondingauthor{Takahiro Morishita}
\email{takahiro@ipac.caltech.edu}

\author[0000-0002-8512-1404]{Takahiro Morishita}
\affiliation{IPAC, California Institute of Technology, MC 314-6, 1200 E. California Boulevard, Pasadena, CA 91125, USA}

\author[0000-0001-9935-6047]{Massimo Stiavelli}
\affiliation{Space Telescope Science Institute, 3700 San Martin Drive, Baltimore, MD 21218, USA}

\author[0000-0003-2497-6334]{Stefan Schuldt}
\affiliation{Dipartimento di Fisica, Università degli Studi di Milano, Via Celoria 16, I-20133 Milano, Italy}
\affiliation{INAF - IASF Milano, via A. Corti 12, I-20133 Milano, Italy}

\author[0000-0002-5926-7143]{Claudio Grillo}
\affiliation{Dipartimento di Fisica, Università degli Studi di Milano, Via Celoria 16, I-20133 Milano, Italy}
\affiliation{INAF - IASF Milano, via A. Corti 12, I-20133 Milano, Italy}




\begin{abstract}
We report JWST/NIRSpec observations of a star-forming galaxy at $z=2.76$, \id. 
We securely detect two temperature-sensitive auroral lines, \siii\,6312 (7.4\,$\sigma$) and \oii\,{7320+7331} doublets (10\,$\sigma$), and tentatively \nii\,5755 ($2.3\,\sigma$) for the first time in an individual galaxy at $z>1$.
We perform a detailed analysis of its interstellar media (ISM), and derive electron temperatures, various heavy element abundances (O/H, N/O, S/O, and Ar/O) in the hot ionized region, and the neutral fraction in the warm ionized region. \id\ shows a broad feature at the wavelength of \heii\,4686, which consists of a broad ($\sim1000$\,km/s), blue-shifted ($\sim-110$\,km/s) line component. Taken together with its mildly elevated N/O abundance, we conclude that \id\ is experiencing a young starburst ($\simlt10$\,Myr), likely hosting a large number of Wolf-Rayet stars. 
None of its spectral features support the presence of AGN, including: $(i)$ the absence of broad components and velocity shifts in Hydrogen recombination lines, $(ii)$ low \feii${\rm 1.257\,\mu m}$/\pb\ ratio, and $(iii)$ the absence of high-ionization lines.
Our analysis using \hei\ lines reveals a higher electron temperature and a higher attenuation value, indicating that \hei\ may probe a smaller spatial scale than \hi, presumably the region dominated by the aforementioned Wolf-Rayet stars. The star formation rates derived from various \hei\ lines broadly agree with those from Hydrogen recombination lines. We thus advocate that \hei\ can be an excellent, independent probe of multi-phase ISM in the era of JWST. 
\end{abstract}



\section{Introduction} \label{sec:intro}
The gas cycle within galaxy systems is fundamental to their formation processes. The interstellar medium (ISM) undergoes continuous chemical enrichment through the infusion of ejecta from previous generations of stars, while simultaneously being diluted by relatively pristine gas infalling from surrounding circumgalactic resources. The accumulation of stars formed through this complex interplay of gas reservoirs shapes the major structures of galaxies observed in later epochs \citep[e.g.,][]{lilly13,angles17}. Therefore, it is crucial to observationally constrain the properties of the ISM in galaxies during their active formation epochs.



The two most important quantities characterizing ISM are the electron temperature ($T_e$) and electron density ($n_e$). These properties not only define the basic nature of ISM but also play a crucial role in determining other properties, such as heavy element abundances. Traditionally, the former has been determined by using an optical auroral line sensitive to temperature, such as \oiiif, whereas the latter is derived from the ratio of doublet lines, such as \siiin\ and \oii\,3726+3729 \citep[e.g.,][\citealt{izotov06} for a detailed analyis]{osterbrock06}. 

Previously, accessing auroral lines in galaxies at $z>1$ posed a significant challenge due to their faintness compared to the bright emission lines of the same species (e.g., $\oiiif/\oiii\,5007\sim1/100$). Despite a few exceptional cases \citep{sanders16,sanders20,gburek19}, these lines have generally remained elusive. However, with the advent of the James Webb Space Telescope (JWST), this situation has changed dramatically. Multiple sources have now been identified with significant \oiiif\ detection, up to redshift of $z=9.5$ \citep{arellano22,schaerer22,nakajima23,sanders23c,curti23,laseter23,morishita24}. Additionally, in some instances, auroral lines other than \oiiif\ have been observed. For example, \citet{sanders23b} reported the detection of \oii\,7320+7331 ($4.4\,\sigma$) in two galaxies at $z=2.18$. Later, \citet{rogers23} reported the same doublet lines, along with another auroral line \siii\,6312 detected at $9\,\sigma$ in Q2343-D40, a galaxy at $z=2.96$. \citet{strom23} showcases a number of faint lines in composite spectrum of galaxies at $z=1$--3, including the auroral \nii\,5755 line. 


In the literature, it has been common practice to use a single temperature to infer multiple element abundances. However, this approach is only valid under certain conditions, namely when emission lines of different species coincide in the roughly same, uniform ionization structure. This assumption often proves inaccurate, particularly at high redshifts due to a number of mergers and other dynamically violent processes \citep[e.g.,][]{tacconi10,genzel11}. Given the capability of JWST enabling multiple auroral line detections for individual systems, ISM studies stand to benefit significantly from more detailed analysis.

\begin{figure*}
\centering
    \includegraphics[width=0.99\textwidth]{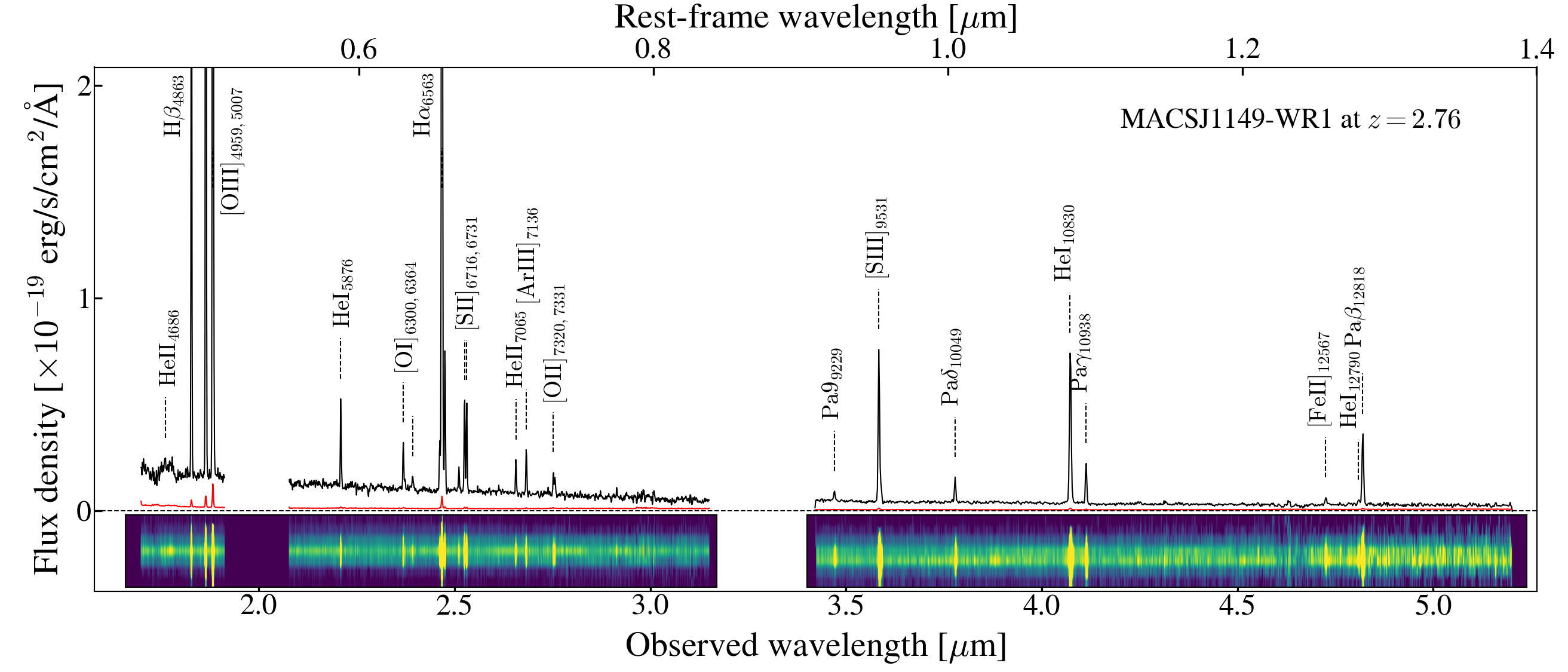}
    \includegraphics[width=0.48\textwidth]{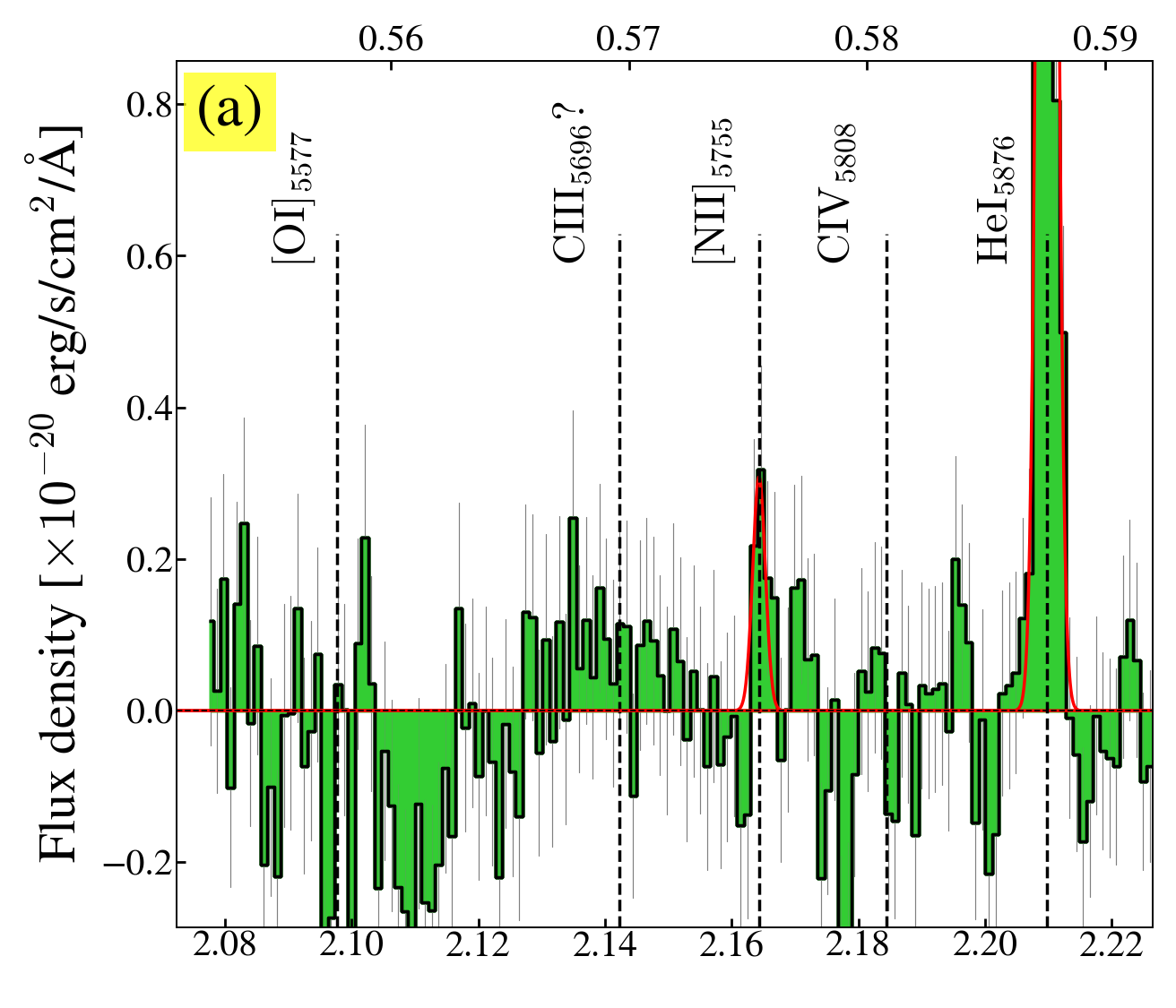}
    \includegraphics[width=0.48\textwidth]{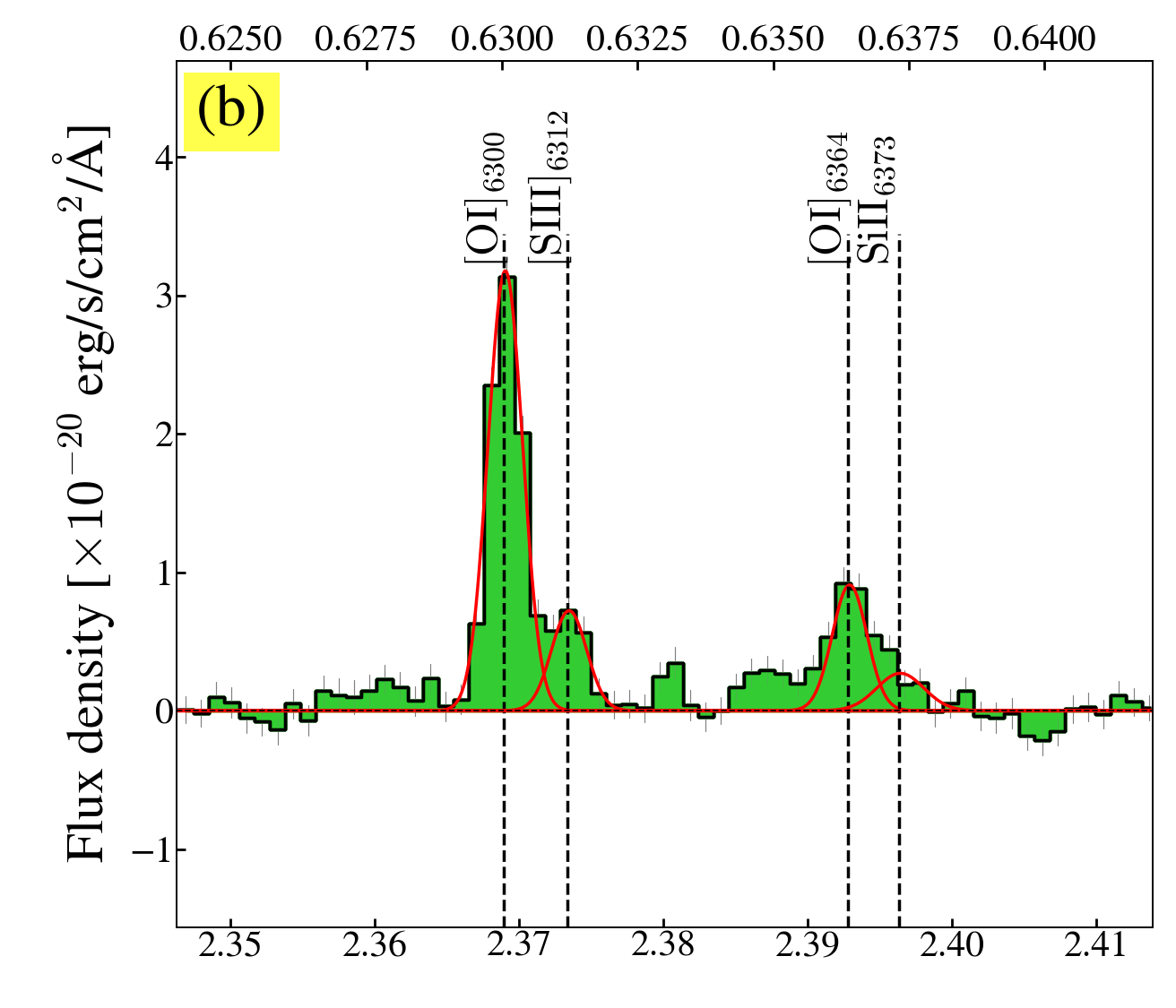}
    \includegraphics[width=0.48\textwidth]{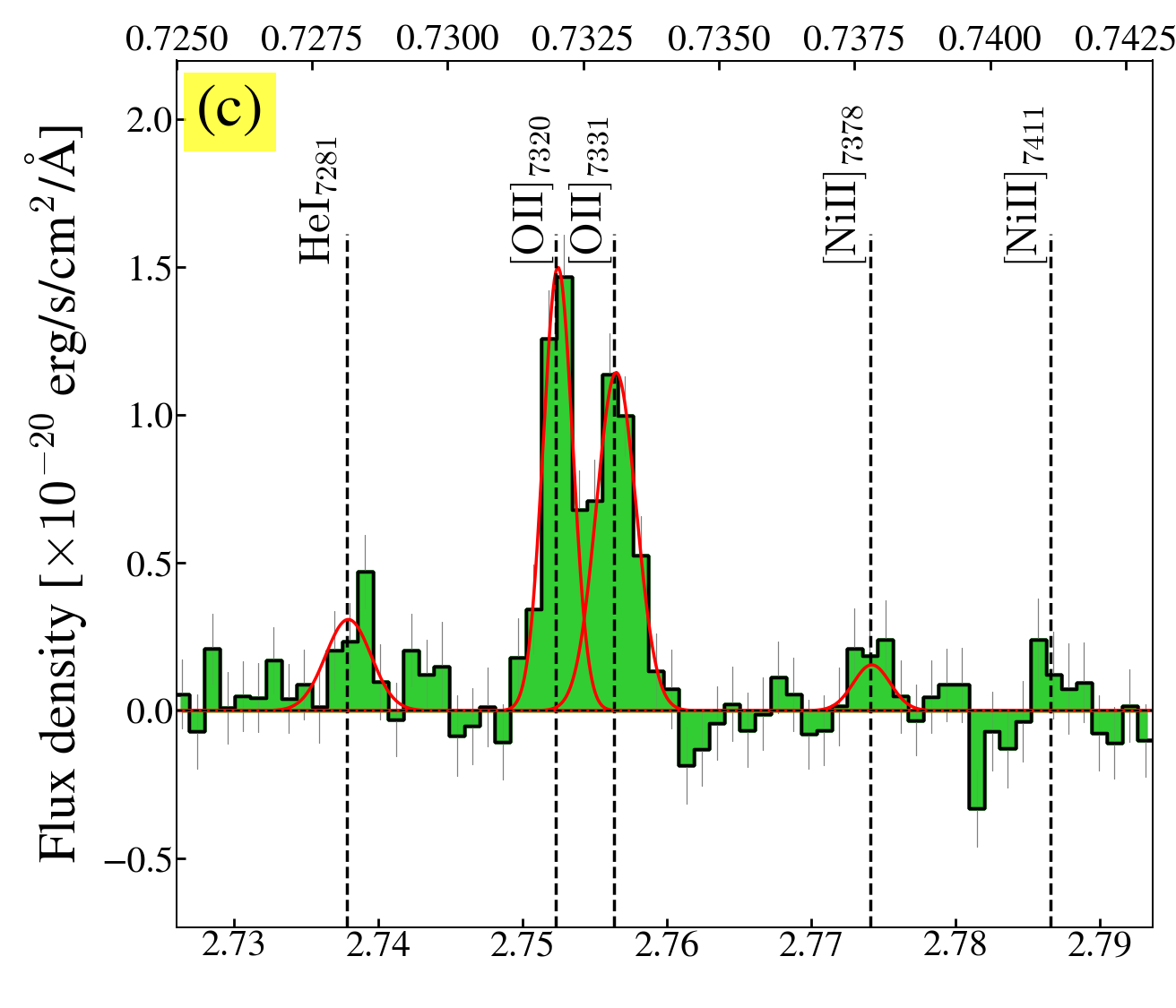}
    \includegraphics[width=0.48\textwidth]{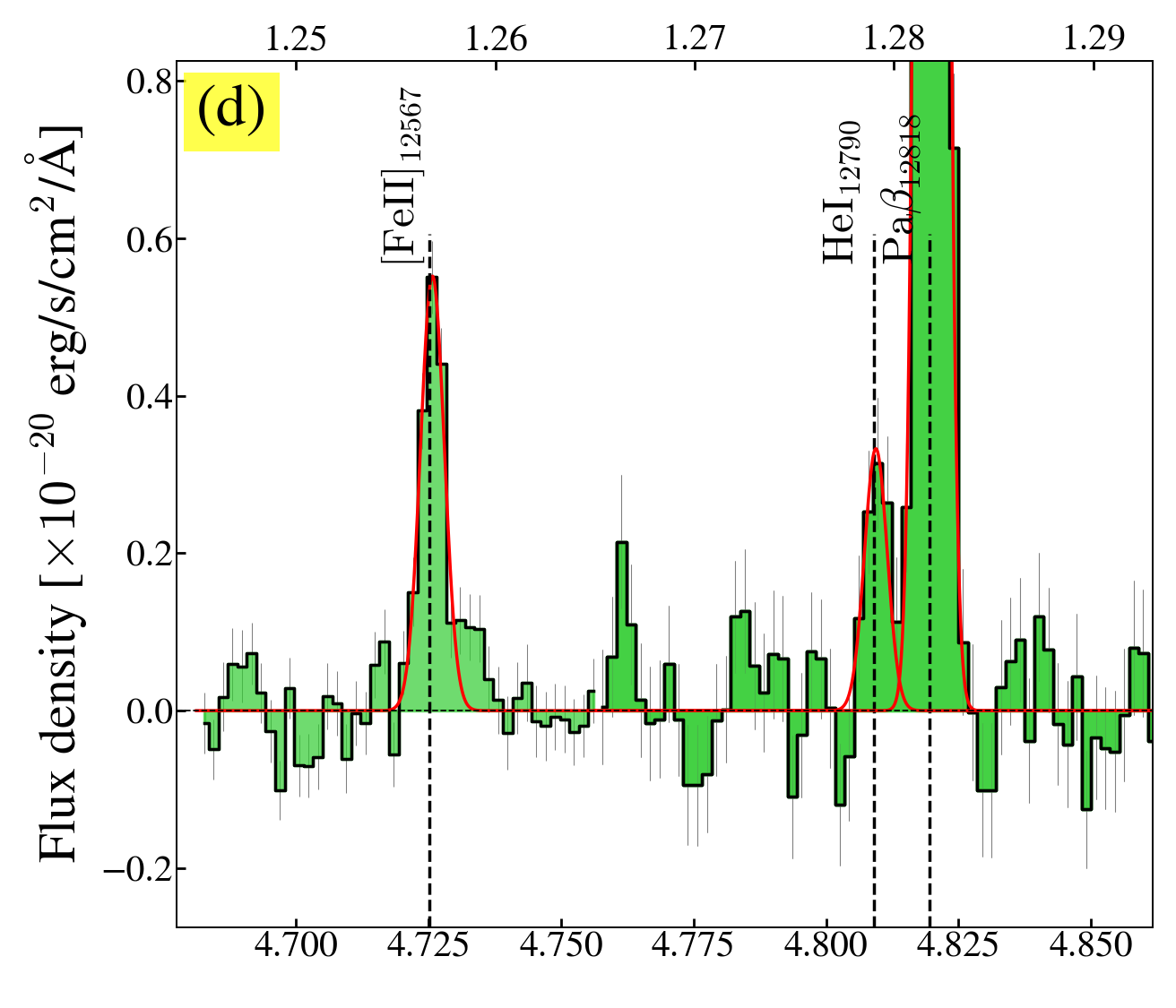}
	\caption{
 (Top): One-dimensional spectrum of \id. Bright lines are labelled. The noise level is also shown (red line). Two-dimensional spectrum is shown at the bottom of the panel (G235M and G395M). The gap within the G235M spectrum is caused by a detector gap.
 (Bottom): Zoom-in plot of the continuum-subtracted spectrum around faint emission lines: (a) \nii\,5755, (b) \siii\,6312, (c) \oii\,7320+7331, and (d) \feii\,1.257\,$\mu$m. The fitted gaussian for each detected line is shown (red lines). The location of \civf\,5808, a characteristic line of WR stars, and \oi\,5755, another auroral line, are indicated despite being undetected. For the \heii\,4686 line complex, see Fig.~\ref{fig:HeII_spectra}.
 }
\label{fig:spectra}
\end{figure*}

In this paper, we report the detection of three auroral lines in a galaxy at $z=2.76$, enabling us to compare electron temperatures derived for different elements. Our deep JWST/NIRSpec spectrum captures \nline\ emission lines spanning the rest-frame range of 0.45--1.3\,$\mu$m (Fig.~\ref{fig:spectra}), facilitating a comprehensive analysis of its ISM properties. Notably, our spectrum also reveals a broad line complex at the wavelength of \heii\,4686, a characteristic feature of Wolf-Rayet stars. 

This paper is structured as follows: In Sec.~\ref{sec:data}, we present our observations and their reduction. In Sec.~\ref{sec:ana}, we perform a spectroscopic analysis, followed by a photometric analysis of the host galaxy in Sec.~\ref{sec:ana_gal}. In Sec.~\ref{sec:NO}, we discuss the inferred element abundances in the context of massive stars. In Sec.~\ref{sec:agn}, we infer the absence of AGN using multiple emission line diagnostics. In Sec.~\ref{sec:he}, we attempt to infer the ISM properties by using a series of \hei\ lines detected across the wavelength range, as an independent probe of ISM. Where relevant, we adopt the AB magnitude system \citep{oke83,fukugita96}, cosmological parameters of $\Omega_m=0.3$, $\Omega_\Lambda=0.7$, $H_0=70\,\kms\, {\rm Mpc}^{-1}$, and the \citet{chabrier03} initial mass function. 

For convenience, we express flux ratios as follows:
$$ R_3 = \log ( (\oiii\,{4959}+\oiii\,{5007})/{\rm H}\beta) $$
$$ R_{2, 7325}=\log ((\oii\,{7320}+\oii\,{7331})/{\rm H}\beta)$$
$$ R_{S2}=\log ((\sii\,{6716}+\sii\,{6731})/{\rm H}\beta)$$
$$ R_{S3}=\log (\siii\,{6312}/{\rm H}\beta)$$
$$ R_{Ar3}=\log (\ariii\,{7135}/{\rm H}\beta)$$
$$ R_{N2}=\log ((\nii\,{6548}+\nii\,{6584})/{\rm H}\beta).$$

\begin{figure}
\centering
    \includegraphics[width=0.48\textwidth]{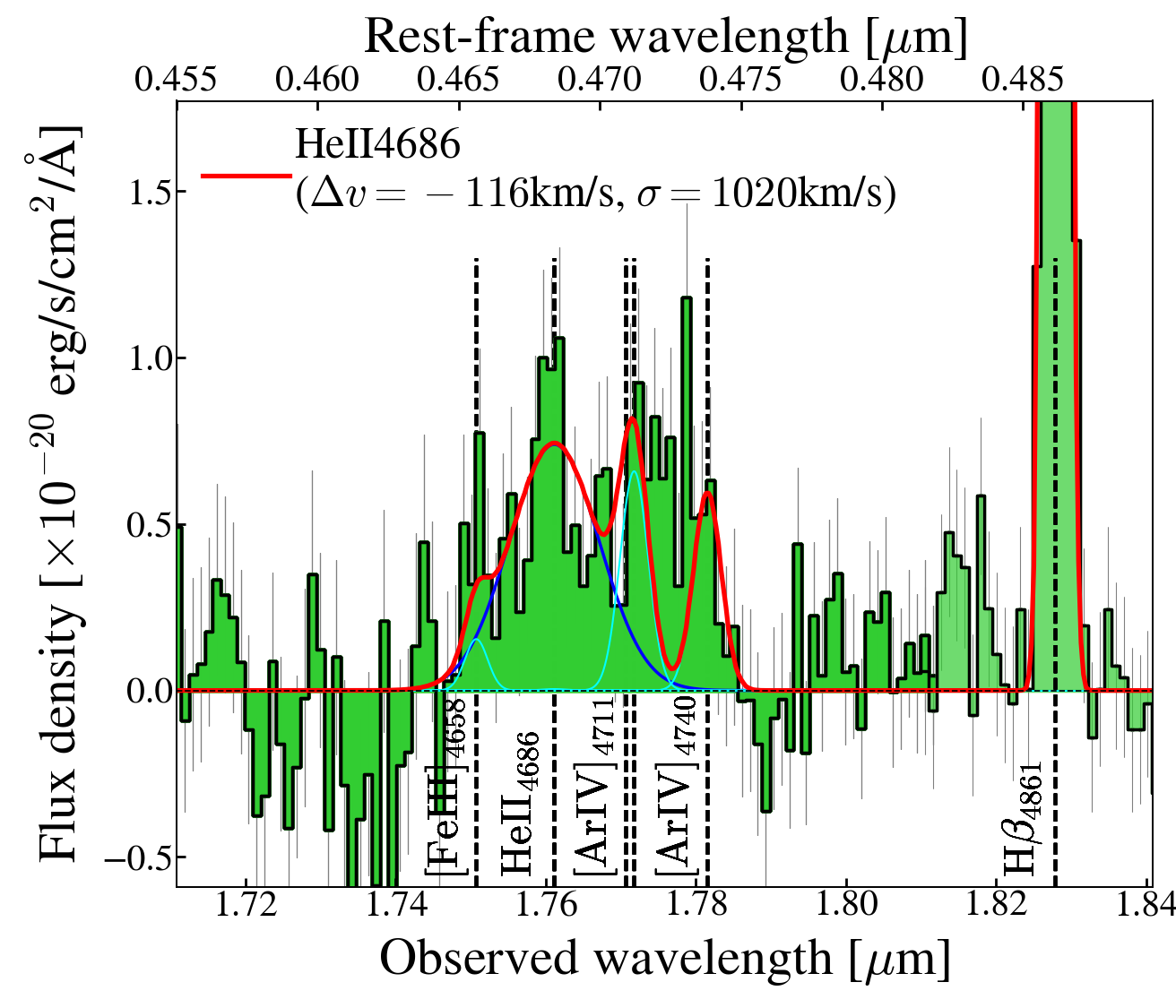}
	\caption{
 Zoom-in plot around the \heii\,4686 complex. The \heii\,4686 complex is fitted with multiple gaussian components: broad \heii\,4686 line (blue line) and narrow \ariv\,4711,4740, and \hei\,4714 lines (thin cyan lines). The total model is shown in red. The \heii\ line is broad ($\sigma\sim1000$\,km/s) and blue-shifted ($\sim-110$\,km/s) from the systemic redshift.
 }
\label{fig:HeII_spectra}
\end{figure}

\section{JWST/NIRSpec and NIRCam Observations}\label{sec:data}

We briefly describe our JWST/NIRSpec and NIRCam observations, conducted as part of the GTO1199 program, targetted on the field of the galaxy cluster MACS~J1149.6+2223 ($z=0.54$; \citealt{stiavelli23}). Readers are referred to \citet{morishita24} for the full details about our observations and data reduction. 


Our target galaxy in this paper, \id, was observed for $9.2$\,hrs (excluding overhead), with each of the G235M and G395M gratings, under the NIRSpec Multi-Object Spectroscopy (MSA) mode. We reduced the MSA data using {\tt msaexp}\footnote{https://github.com/gbrammer/msaexp} ({ver.~0.8.5}), by following \citet{morishita23b}. The one-dimensional spectrum was extracted via optimal extraction. 


We followed the same steps for JWST/NIRCam imaging reduction and photometry as presented in \citet{morishita23}. 
We identified sources in the detection image using SExtractor \citep{bertin96}, and measured their fluxes on the psf-matched mosaic images with a fixed aperture of radius $r=0.\!''16$\,. \id\ is located in an area relatively free from contamination by surrounding sources and foreground cluster galaxies, assuring a robust flux estimate. The magnification by the foreground cluster is estimated to be $\mu = 1.34_{-0.01}^{ +0.02}$ (Schuldt et al., in prep.). While the magnification factor is not as significant as for sources near the cluster center, we note that this effectively increases the on-source exposure time by a factor of $\mu^2 \sim 1.8$, thereby significantly improving S/Ns.





\section{ISM Analysis}\label{sec:ana}
In what follows, we describe our analysis of the ISM using the extracted 1d spectrum.

\begin{figure}
\centering
    \includegraphics[width=0.48\textwidth]{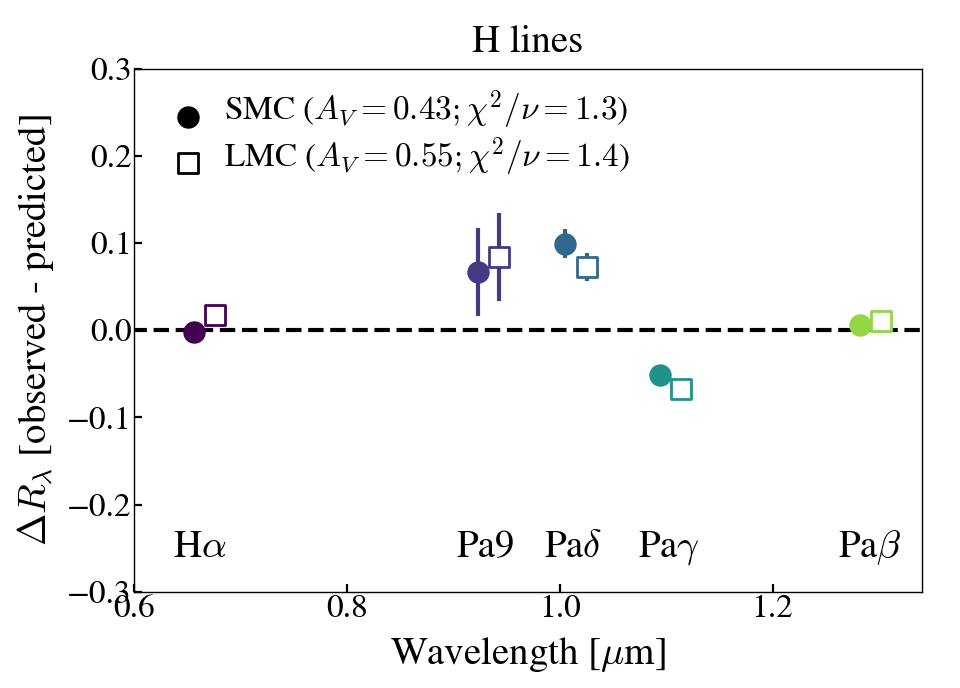}
	\caption{
 Residual of the line flux extinction correction, $\Delta R_\lambda = (I_{\lambda}/I_{\rm H\beta})_{\rm measured} - (I_{\lambda}/I_{\rm H\beta})_{\rm model}$, for various Hydrogen recombination lines. The results with the SMC (LMC) dust model are shown with filled circles (empty squares). For \id, the SMC dust model provides a slightly better fitting result. \hb\ is not shown, as the line is used as the pivotal wavelength (i.e. $\Delta R_\lambda = 0$).
    }
\label{fig:attn}
\end{figure}

\subsection{Emission Line Measurement}\label{linefit}

We model each emission line of interest with a Gaussian function. Our basic strategy is to define a wavelength window for each line, model the underlying continuum spectrum by a 2nd-order polynomial fit, and subtract it from the 1-d spectrum before the Gaussian fit is carried out. The model includes three parameters, amplitude, line width, and redshift. When multiple lines are located close to each other, we fit those simultaneously using a common redshift parameter, so blending can effectively be resolved. In addition, doublet lines (i.e. \oiii$_{\lambda\lambda4959+5007}$ and \nii$_{\lambda\lambda6550+6585}$) are modelled with a fixed line ratio (1:3 and 1:1.7, respectively), with single line width and redshift parameters. The total flux of the doublets estimated in this way remains robust, while we note that the ratio could have a systematic uncertainty for $\lesssim10\,\%$ \citep[e.g.,][]{storey00}. {For doublet lines that we aim to measure the ratio (i.e. \oii\,7320,7331 and \sii\,6718,6733), we fit them without having such constraints.}

The integrated flux of each line is estimated by summing the corresponding Gaussian component, and the flux error is estimated by summing the error weighted by the amplitude of the Gaussian model in quadrature. {We then compare this flux error with the one originating in the model uncertainty and take the one larger for each line. By having these two uncertainty estimates, the final flux error is conservative and accurate in potential anomalous occasions, e.g., when one of the doublet lines fall on a detector gap.} In the following analysis, we adopt flux measurements when the line is detected (signal-to-noise ratio ${\rm S/N}\geq2$); for those not detected, we adopt 2-$\sigma$ flux upper limits. 
We detect \nline\ lines at S/N\,$>2$. We report emission line fluxes estimated in Table~\ref{tab:lines}. 

\subsubsection{Faint Auroral Lines}\label{sec:auroral}
We detect auroral emission lines, \siii\,6312 and \oii\,7320,7331 at a significant confidence level ($7.1$\,$\sigma$, $10.8$\,$\sigma$, 9.3\,$\sigma$, respectively), and tentatively \nii\,5755 line ($2.4\,\sigma$; Panels (a)--(c) in Fig.~\ref{fig:spectra}). {The \nii\ line was successfully fit with a Gaussian but the resulting width was found to be narrower than the resolution limit. As such, we fix the line width to the one defined by spectral resolution at the corresponding wavelength ($R\sim910$, or $\sim140$\,km/s), making its flux estimate conservative. We note that widths of other lines are also found similarly narrow, which supports our procedure on \nii\ here.}

These lines are used to infer the electron temperature (Sec.~\ref{sec:te}), which is essential for robust inference of element abundances (Sec.~\ref{sec:metal}). 
The ratio of the \oii\ doublets is $I_{7320}/I_{7331}=0.97\pm0.12$. 
While the \siii\,6312 line is located near \oi\,6300 and partially blended, the peak of the line is clearly separated and enables robust flux estimate. 

\subsubsection{Broad \heii\,4686 Line Complex}\label{sec:heii}
We find a broad feature at the wavelength of \heii\,4686 (Fig.~\ref{fig:HeII_spectra}). The feature is characteristic to galaxies dominated by Wolf-Rayet (WR) stars \citep[e.g.,][]{allen76,brinchmann08}, often referred to as blue bump. In addition to broad \heii\,4686, WR galaxies often show broad \niiif\,4640 and \ciiif\,4650 emission lines, {depending on the type of Wolf-Rayet stars (Carbon-dominated WC or Nitrogen dominated WN) and their sub-classes \citep{smith68,hucht01}}, as well as narrow lines such as \feiii\,4658, \ariv\,4711,4740, and \hei\,4714. 

We attempt to fit the line complex with a multi-component gaussian model. After experimenting various combinations, we found that a simple model that consists of one broad component (corresponding to \heii) and four narrow components (\feiii, \ariv\ doublets, \hei) well reproduces the complex within reasonable parameter ranges. The redshift of all components is set to a single value during the fit, to avoid extra complexity. 

With this configuration, we find that the broad \heii\ component is characterized broad ($\sigma\sim1000$\,km/s) and blue-shifted for $\sim-110$\,km/s from the systemic redshift. The derived line width is large but within the range of local WR galaxy samples \citep[e.g.,][]{brinchmann08}.

The blue bump spectral feature can be characterized by its flux ratio to, e.g., \hb. We find this to be $\sim0.15$ for \id, which is consistent with the range found in local galaxies \citep[e.g., $\sim 0.1$--$1$;][]{sargent91}.

The addition of broad \niiif\ and/or \ciiif\ component turned out unnecessary in our case. Those two lines fall near the blue edge of the observed emission feature. In fact, the model including those was found incompatible without introducing an unreasonably large velocity shift ($\simgt +2000$\,km/s). 
We do not clearly see \ciiif\,5696 and \civf\,5808 in our spectrum, i.e. another characteristic feature called ``red bump," though we notice there exists slight excess of flux at the corresponding wavelength  (Fig.~\ref{fig:spectra}). As such, the absence of significant \ciiif\ and \civf\ emission near the \heii\ complex is not surprising. 

Due to the sensitivity limitation in our spectrum, further characterization of the underlying WR stars should defer to a future study with additional data. Yet, our analysis has clearly characterized \id\ as a galaxy likely hosting a non-negligible number of WR stars (see also Sec~\ref{sec:NO}).


\subsubsection{Other Emission Lines}\label{sec:other}
Besides the lines remarked in the previous two subsections, we note the detection of lines that underrepresent in ISM studies in the literature, including \silii\,6373 (Fig.\ref{fig:spectra}, panel b), \feii\,1.257\,$\mu$m (d), and a series of \hei\ lines. While \feii\ has a low ionizaion potential (7.9\,eV), it is often seen in AGN or shock-dominated regions. This is because iron element is highly depleted on dust grains and the element is observable upon the destruction of dust grains. As such, a high \feii\ line ratio (e.g., to Pa$\beta$) would indicate the presence of such mechanisms in act (see Sec.~\ref{sec:agn}). For the same reason, \feii\ is not a good line for iron abundance estimate, as the fraction of iron atoms in the ISM over those depleted on dust grains is not easily observable. 

Overall, all the detected lines are found to be narrow, with $\sigma\sim100$--200\,km/s, comparable to the instrumental resolution, except for those in the aforementioned \heii\ line complex. Some Hydrogen recombination lines are detected at high significance, and yet we do not observe clear evidence of a broad component.

\subsection{Dust Extinction}\label{sec:ext}
Extinction in ISM is estimated by using the detected Hydrogen recombination lines i.e. \hb, \ha, Pa9, \pd, \pg, and \pb. The intrinsic emissivity of each line is estimated under the assumption of Case~B, with $T=10000$\,K \citep{osterbrock06}. While the recombination coefficients have temperature dependence, we find in the following section that our assumption is reasonable. We estimate the attenuation, $A_V$, via chi-square minimization between the observed and the predicted line ratios in the base of \hb, $R_\lambda=I_{\rm \lambda}/I_{\rm H\beta}$. Fig.~\ref{fig:attn} shows the comparison, where we find $A_V=0.43$ for the SMC dust curve \citep{gordon03} and $A_V=0.55$ for the LMC curve \citep{calzetti00}. We find that the SMC model gives a slightly better fitting result ($\chi^2/\nu=1.3$) over the LMC's ($=1.4$). As such, in the following analysis, we adopt the correction obtained with the SMC curve.

\begin{figure*}
\centering
    \includegraphics[width=0.8\textwidth]{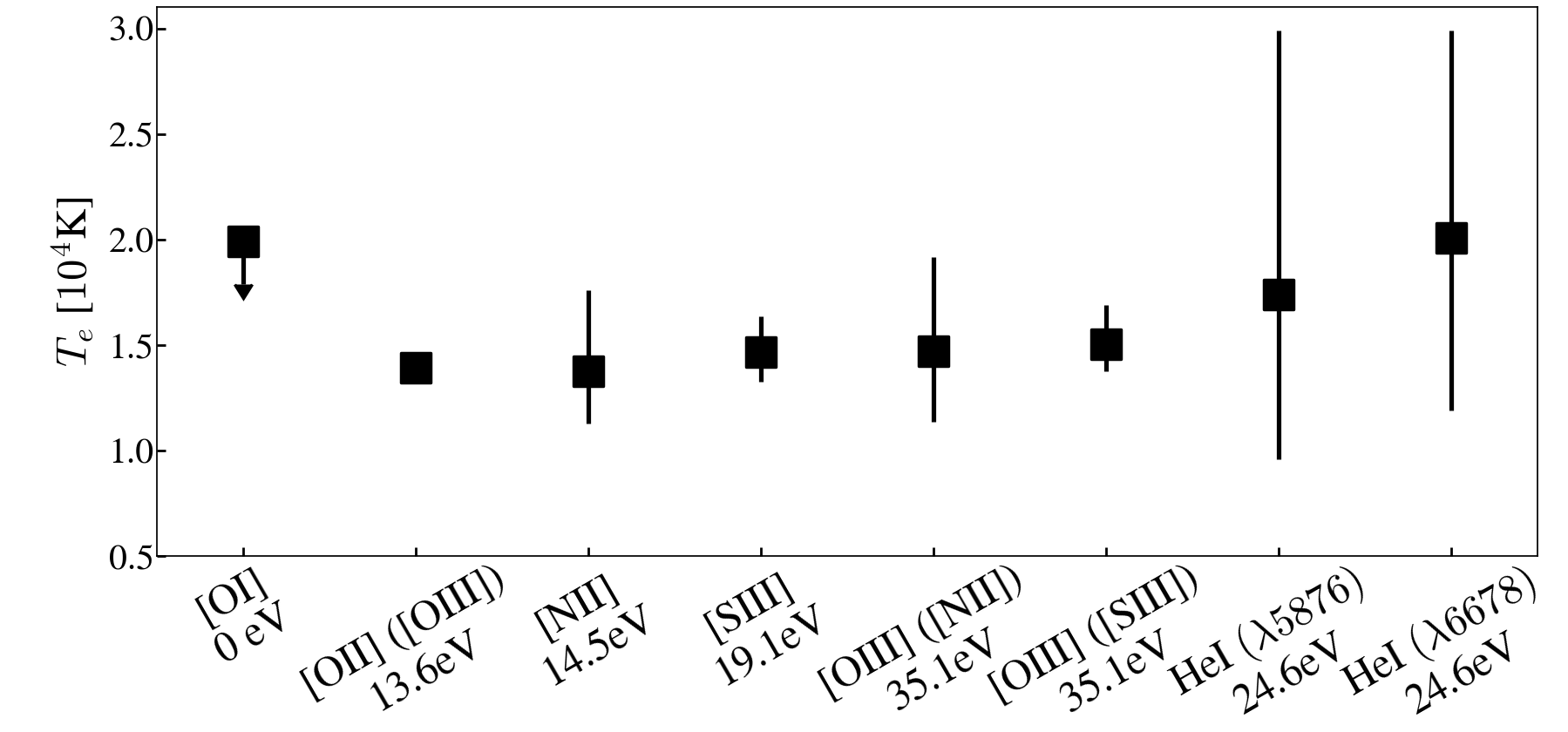}
	\caption{
 Electron temperatures of \id\ measured with various temperature sensitive lines (Sec.~\ref{sec:te}). Ionization potential is shown in the label of each line. $\oiii\ (\siii)$ is the temperature derived from $T_e(\siii)$ (Eq.\ref{eq:toiiisiii}) and $\oiii\ (\nii)$ from $T_e(\nii)$ (Eq.~\ref{eq:toii}). $T_e(\oi)$ is derived using the lower limit \oi\,(6300+6364)/\oi\,5577, and thus the derived $1\,\sigma$ upper limit is shown. 
 Two temperature measurements, $T_e(\hei)$, derived using the line ratios \hei\,7281/\hei\,5876 and \hei\,7281/\hei\,6678, are shown (Sec.~\ref{sec:he}).
    }
\label{fig:tes}
\end{figure*}

\subsection{Electron Density and Temperature} 
\label{sec:te}
Electron density can be estimated from doublet lines such as \oii\,3726,3729 and \sii\,6718,6733. However, electron density has a mild dependency on electron temperature, and vice versa. As such, it is desirable to determine both quantities simultaneously, e.g., through an iterative approach. 

We first determine electron density and electron temperature using two line ratios, \sii\,6718/\sii\,6733 and \siii\,(9531+9069)/\siii\,6312. For electron density, we use the formula presented in \citet{sanders20}, with the temperature dependence as outlined in \citet{osterbrock06}:
\begin{equation}\label{eq:1}
n_e = \frac{(c\,\sii_{6716}/ \sii_{6731} - a\,b)}{(a - \sii_{6716}/ \sii_{6731})} (\frac{T_e}{10^4})^{-1/2},
\end{equation}
with $a=0.4315$, $b=2107$, and $c=627.1$. For electron temperature, we use Eq.~(5.7) in \citet{osterbrock06}:
\begin{equation}
\frac{\siii_{9069}+\siii_{9531}}{\siii_{6312}} = \frac{5.44 \exp(2.28 \times 10^4/T_e)}{1 + 3.5\times10^{-4} n_e / T_e^{1/2}},
\end{equation}
{where $n_e$ and $T_e$ hold the same values as in Eq.~\ref{eq:1}}.
Note that the \siii\,9096 line falls in the detector gap, and thus we instead use a fixed ratio of the doublets, $I_{9531}/I_{9096} = 2.5$ \citep{sanders20}. By solving the two equations above simultaneously, we find {$n_e(\sii)=363_{
-64}^{+76}$\,cm$^{-3}$ and $T_e(\siii)=1.47_{-0.14}^{+0.17}$\,K}. Our spectra do not have the wavelength coverage for the \oii\,3726,3729 doublets. The \oii\,7320,7331 doublet ratio has almost no dependency on $n_e$. Therefore, we adopt $n_e(\sii)$ in the following analysis. {Throughout the following analyses, we propagate the uncertainty of $n_e$, in addition to the uncertainties of flux measurements of interest.}

{Since} our spectra do not cover the temperature sensitive $\sii_{4068,4076}$ lines \citep[e.g.,][]{shaw95}, we cannot directly obtain $T_e(\sii)$.
{
However, our iterative analysis above implicitly assumes $T_e(\sii)=T_e(\siii)$, whereas in the local universe the former is found systematically lower \citep[$\sim$\,a few $1000$\,K, e.g., ][]{perez-montero17}. (Also, see below where we derive the temperature from ${\rm N^+}$, which has a similar ionizing potential to ${\rm S^+}$.) However, we find the impact from this remains relatively small. For example, by setting $T_e(\sii)=T_e(\siii)-2000$, we find $n_e=338\pm47$\,cm$^{-3}$. The change in $n_e$ has little impact in the following analyses. As such, we adopt the $n_e$ measurement obtained with the $T_e(\sii)=T_e(\siii)$ assumption in the following analysis.}

With \nii\,5755, we use Eq.~(5.5) in \citet{osterbrock06}:
\begin{equation}
\frac{\nii_{6548}+\nii_{6583}}{\nii_{5755}} = \frac{8.23 \exp(2.50 \times 10^4/T_e)}{1 + 4.4\times10^{-3} n_e / T_e^{1/2}},
\end{equation}
and obtain $T_e({\rm \nii})=1.37_{-0.25}^{+0.38}\times10^4$\,K. The larger error in the estimate here comes from the measurement uncertainty in the \nii\,5755 flux.

We then attempt to estimate the temperatures of \oii\ and \oiii, which are critical to oxygen abundance estimate. This is, however, tricky in our case, since our spectra do not cover the wavelength range of \oii\,3726,3729 nor \oiii\,4363. As such, in the following analysis we assume $T_e(\oii)=T_e(\nii)$, as the two lines have similar ionization potential (13.6\,eV and 14.5\,eV). 

$T_e(\oiii)$ can be inferred from $T_e(\siii)$, by using Eq.(15) in \citet{izotov06}:
\begin{equation}\label{eq:toiiisiii}
\begin{split}
t_e (\siii) & = -1.276 \\ & + t_e (\oiii) (2.645 - 0.546 t_e (\oiii)),
\end{split}
\end{equation}
where $t_e$ represents the corresponding temperature scaled by $1/10^4$. By solving the equation, we obtain $T_e(\oiii)_{\siii}=1.50_{-0.13}^{+0.19}\times10^4$\,K. 

Alternatively, one can estimate $T_e(\oiii)$ via $T_e(\oii)$ by using Eq.(14) in \citet{izotov06}:
\begin{equation}\label{eq:toii}
\begin{split}
t_e (\oii) & = -0.744 \\ & + t_e (\oiii) (2.338 - 0.610 t_e (\oiii)).
\end{split}
\end{equation}
This gives $T_e(\oiii)_{\nii}=1.47_{-0.33}^{+0.45}\times10^4$\,K, which is consistent with $T_e(\oiii)_{\siii}$ but has a larger uncertainty. We adopt $T_e(\oiii)_{\siii}$ for the oxygen abundance inference in the following section. 

{As a sanity check, we derive $T_e(\oii)$ via Eq.~\ref{eq:toii}, by using $T_e(\oiii)_{\rm \siii}$. This gives $T_e(\oii)=1.4\pm0.1\times10^4$\,K. The estimate is consistent with $T_e(\nii)$, which validates the assumption, $T_e(\oii)=T_e(\nii)$, made above.} 

Lastly, with the lower limit flux ratio \oi\,(6300+6364)/\oi\,5577, we find $T_e(\oi)<1.99 \times 10^4$\,K ($1\,\sigma$).

For Ar$^{2+}$, while we detect \ariii\,7135 and 7751 lines, our spectrum do not have the wavelength coverage for \ariii\,5192, due to the detector gap, which can be a good temperature indicator by combining with the doublet lines \citep[e.g.,][]{keenan88,proxauf14}. We thus adopt $T_e(\siii)$ in the Ar$^{2+}$ abundance inference below.

In Fig.~\ref{fig:tes}, we compare the derived temperatures. Overall, they are in good agreement within the uncertainties. We do not see any significant trend in $T_e$ with ionizing potential, indicating that their ionizing regions likely coincide.

\subsection{Heavy Element Abundances} 
\label{sec:metal}
Now equipped with the electron density from \sii\ and electron temperature from multiple lines, we attempt to obtain heavy element abundances. 

Oxygen abundance is estimated by deriving O$^{2+}$/H$^{+}$ and O$^{+}$/H$^{+}$ separately. Each abundance can be estimated by substituting the temperatures into the following two equations from \citet{izotov06}:
\begin{equation}\label{eq:6}
\begin{split}
    12 + \log & {\rm O^{2+}/H^+} = R_3 + 6.200 + 1.251 / t_e(\oiii) \\ & - 0.55 \log t_e(\oiii) - 0.014 t_e(\oiii),
\end{split}
\end{equation}
\begin{equation}\label{eq:7}
\begin{split}
    12 + \log & {\rm O^+/H^+} = R_{2, 7325} + 6.901 + 2.487 / t_e(\oii) \\ & - 0.483 \log t_e(\oii) - 0.013 t \\ & + \log (1 - 3.48 x),
\end{split}
\end{equation}
where $x = 10^{-4} n_e t_e^{-0.5}$. We obtain
$12 + \log {\rm O^{2+}/H^+} = 7.78_{-0.10}^{+0.08}$ and $12 + \log {\rm O^{+}/H^+} = 7.34_{-0.04}^{+0.04}$. The contribution from the atomic oxygen and triply ionized oxygen are considered negligible. As a sanity check, we run cloudy to calibrate \oiii\,5007/\oiii\,7320 to ionization parameter $\log U$. For the observed ratio, we find $\log U\sim-2.6$, where $O^0$ is expected to be $<10\,\%$ and $O^{3+}\ll1\,\%$ relative to the total oxygen abundance \citep[e.g.,][]{berg19}.
As such, we adopt 
\begin{equation}
{\rm \frac{O}{H} = \frac{O^+}{H^+} + \frac{O^{2+}}{H^+}}
\end{equation}
and obtain $\logoh = 7.92_{-0.08}^{+0.07}$.

We follow similar approaches for Sulfur and Nitrogen using the following equations from \citet{izotov06}: 
\begin{equation}
\begin{split}
    12 + \log & {\rm S^{+}/H^+} = R_{S2} + 5.439 + 0.929 / t_e(\sii) \\ & - 0.28 \log(t_e(\sii)) - 0.018 t_e(\sii) \\ & + \log(1 + 1.39 x),
\end{split}
\end{equation}
\begin{equation}
\begin{split}
    12 + & \log {\rm S^{2+}/H^+} = R_{S3} + 6.690 + 1.678 / t_e(\siii) \\ & - 0.47 \log(t_e(\siii)) - 0.010 t_e(\siii),
\end{split}
\end{equation}
\begin{equation}
\begin{split}
    12 + \log & {\rm N^{+}/H^+} = R_{N2} + 6.234 + 0.950 / t_e(\nii) \\ & - 0.42 \log t_e(\nii) - 0.027 t_e(\nii) \\ & + \log (1 + 0.116 x),
\end{split}
\end{equation}
and obtain $5.49_{-0.01}^{+0.01}$, $5.98_{-0.05}^{+0.05}$, and $6.31_{-0.23}^{+0.22}$, respectively. 
For nitrogen, we assume 
${\rm \frac{N}{O} \approx \frac{N^+}{O^+}}$
, which gives good approximation within $<10\,\%$ uncertainties \citep[e.g.,][]{nava06,amayo21}, and obtain $\log {\rm (N/O)_{\nii}} = -1.04_{-0.23}^{+0.22}$. Using $T_e(\siii)$ instead, we obtain a slightly smaller N/O abundance, $\log {\rm (N/O)_{\siii}} = -1.09$, but with a smaller error bar. Both measurements are consistent with each other.

For Sulfur, we use ionization correction factor ICF ($1.07$) calculated with Eq.~(20) in \citet{izotov06} and obtain $\log {\rm (S/O)} = -1.76_{-0.04}^{+0.04}$.

For Ar, we do not have direct electron temperature measurements. We instead use $T_e(\siii)$ and follow Eq.(11) in \citet{izotov06}:
\begin{equation}
\begin{split}
    12 + \log & {\rm Ar^{2+}/H^+} = R_{Ar3} + 6.174 + 0.799 / t_e(\siii) \\ & - 0.48 \log t_e(\siii) - 0.013 t_e(\siii) \\ & + \log (1 + 0.220 x),
\end{split}
\end{equation}
With the ICF calculated for ${\rm Ar^{2+}}$ ($=1.10$) {using \citet{izotov06}}, we obtain $\log {\rm (Ar/O)} = -2.40_{-0.02}^{+0.02}$. We note that ${\rm Ar^{3+}}$ abundance is generally smaller than that of ${\rm Ar^{2+}}$. As such, we decided to derive the Ar abundance only using ${\rm Ar^{2+}}$.

\subsection{Inference on Neutral Gas Fraction}\label{sec:neut}
Since the \oi\ emission primarily originates due to collisional excitation by thermal electrons, its intensity is a measure of the neutral hydrogen content within the warm ionized regions {when the oxygen abundance is known}. Using \oi\,6300/\ha\ ratio, one could infer the neutral gas fraction therein. We follow the formula in \citet[][]{reynolds98}:
\begin{equation}
\begin{split}
    \frac{I_{6300}}{I_{\rm H\alpha}} & = 2.63 \times 10^4 \frac{n({\rm H^0})}{n({\rm H^+})} \xi \frac{n({\rm O})}{n({\rm H})} \\ &
    \times \frac{t_e^{1.85}}{(1+0.605\, t_e^{1.105})} \exp{(-\frac{2.284}{t_e})},
\end{split}
\end{equation}
where $n({\rm O})/n({\rm H})$ is the gas-phase abundance of oxygen, and $\xi=(1+r)/(\frac{8}{9}+r)$, where $r=n({\rm H^0})/n({\rm H^+})$. By using the oxygen abundance and electron temperature obtained with \nii, which has among the lowest ionization potential, we find $n({\rm H^+})/n({\rm H^0})\sim250$. Applying $T_e(\siii)$ gives a slightly higher ratio, $\sim300$. 
{The obtained ratio in either case is much higher that those seen in typical warm ionized components at the Galactic plane \citep[$\ll100$;][]{reynolds98,haffner09}. This likely suggests that our spectrum consists of lines from two ionized regions in \id\ i.e. hot and warm. However, this fact should not affect the main analysis of our paper, which concerns the former. For example, by using a typical ratio $n({\rm H^+})/n({\rm H^0})=10$, it is suggested that only $<5\,\%$ of the observed \ha\ flux comes from the warm component. While warm ionized media are typically found diffuse and extended, in our two-dimensional spectrum (Fig.~\ref{fig:spectra}) we do not find evidence for that, i.e. the spatial extent of the \oi\ line is not significantly different from those of other lines. Spatially resolved spectroscopy may offer further details.}

We note that the measured ratio of the doublets, $\oi\,6300/\oi\,6364\sim3.5$, confirms that \oi\ emission originates in an optically thin region. {Smaller ratios ($<3$ i.e. optically-thick case)} are only seen in young Supernova remnants, of $\simlt 1 $\,yr after the outburst \citep[e.g.,][]{li92,elmhamdi11}.


\begin{figure*}
\centering
    \includegraphics[width=0.99\textwidth]{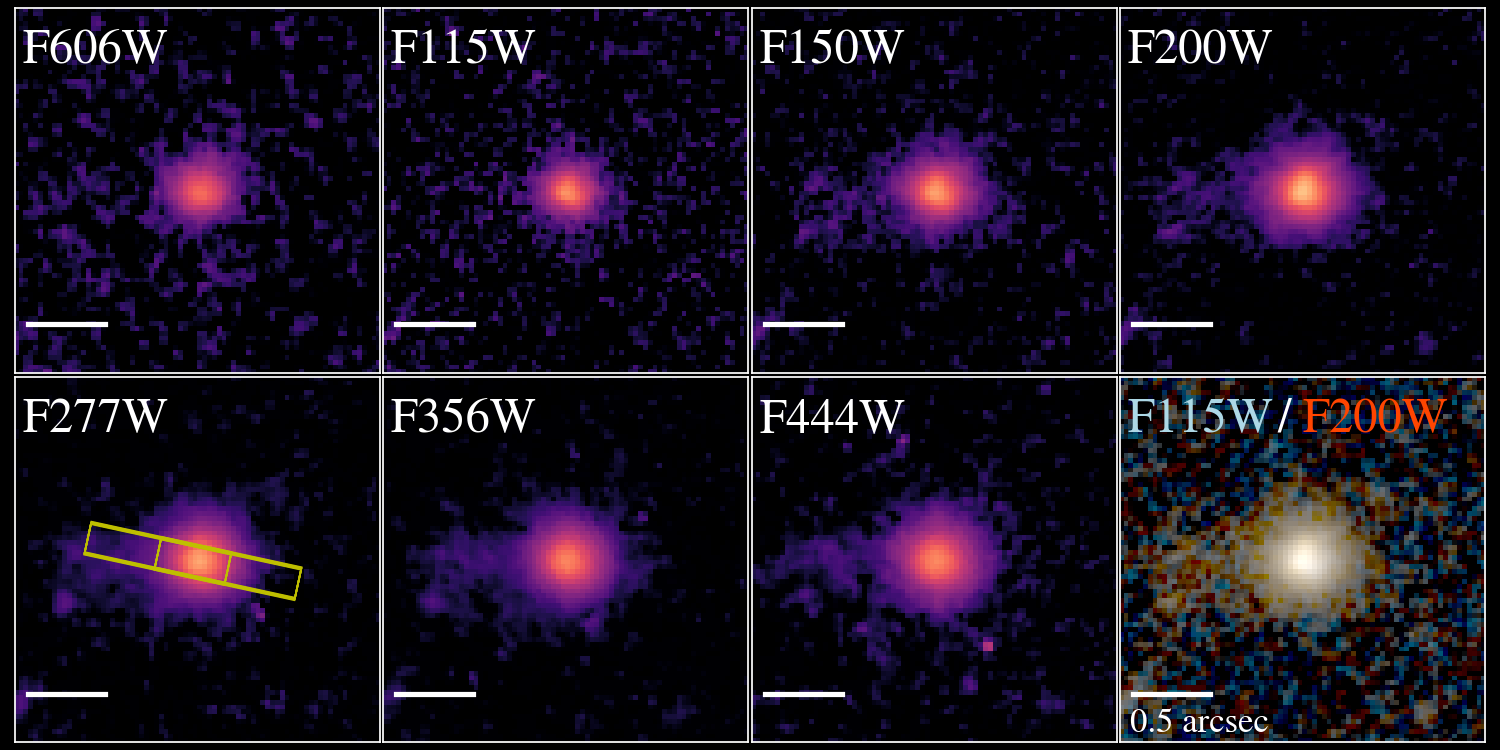}
	\caption{
 Cutout images of \id\ in multiple filters, in the size of $2.\!''4$ ($\sim18$\,kpc). \id\ is characterized by compact morphology across the observed wavelength range. On the F277W stamp, we show the position of our NIRSpec MSA shutters.
    }
\label{fig:stamp}
\end{figure*}

\section{Host Galaxy Analysis}\label{sec:ana_gal}

\subsection{Spectral Energy Distribution Analysis}\label{sec:sed}
We infer the spectral energy distribution (SED) of \id\, by using SED fitting code {\tt gsf} \citep[ver1.85;][]{morishita19}. We utilize the JWST+HST photometric data for the SED analysis. The detailed procedure is {outlined} in \citet{morishita24}. 

The SED-based star-formation rate {(SFR)} is calculated with the rest-frame UV luminosity ($\sim1600\,\rm{\AA}$) using the determined SED model. The UV luminosity is corrected for dust attenuation using {the UV continuum slope $\beta_{\rm UV}$}, which is measured by using the posterior SED, as in \citet{smit16}:
\begin{equation}
    A_{1600} = 4.43 + 1.99\,\beta_{\rm UV}
\end{equation}
The attenuation corrected UV luminosity is then converted to SFR via the relation in  \citet{kennicutt98_sfr}:
\begin{equation}
    {\rm SFR\,[M_\odot\,yr^{-1}]} = 1.4 \times 10^{-28} L_{\rm UV}\,[{\rm erg\,s^{-1}\,Hz^{-1}}].
\end{equation}
{The estimated SFR is then corrected for the Chabrier IMF, by multiplying a factor of 0.63, which is estimated from {\tt fsps} \citep[e.g.,][]{madau14}.}
The measured properties are summarized in Table~\ref{tab:phys}. 

\subsection{Morphology Analysis}\label{sec:lp}
\id\ is characterized by compact morphology (Fig.~\ref{fig:stamp}). We measure the effective radius of \id\ using {\tt galfit}, in the same way as in \citet{morishita23b}. The derived sizes are $R_e=2.9$\,pixel, or $0.7$\,kpc in F444W (rest-frame IR), $0.6$\,kpc in F200W (rest-frame optical), and $0.6$\,kpc in HST/F606W (rest-frame UV). While the measured size is overall small, it is not significantly off from those of star-forming galaxies at similar redshifts \citep[e.g.,][]{vanderwel14}. Using the stellar mass and SFR estimated from our SED analysis, 
we find stellar mass surface density of $\Sigma_*\sim2.6\times10^9\,M_\odot {\rm kpc^{-2}}$ and SFR surface density of $\Sigma_{\rm SFR}\sim25\,M_\odot {\rm yr^{-1}  kpc^{-2}}$. As a reference, compact star forming galaxy populations known as blue nuggets, typically have $\Sigma_*\simgt10^9\,M_\odot {\rm kpc^{-2}}$, and as high as $\sim10^{11}\,M_\odot {\rm kpc^{-2}}$ \citep[e.g.,][]{barro17}. As such, \id\ can be considered as a blue nugget but near the boundary.

\id\ is well characterized with a single exponential disk profile in all the filters above, with axis ratio close to $\sim1$. We do not find any disturbed feature or multiple peaks within the central part of the galaxy, nor does it have strong color gradient in F115\,$-$\,F200W (corresponding to rest-frame Balmer break). However, it has a faint feature extended to the East direction.

\input{table_linefluxes.tex}


\section{Discussion}\label{sec:disc}

\begin{figure}
\centering
    \includegraphics[width=0.49\textwidth]{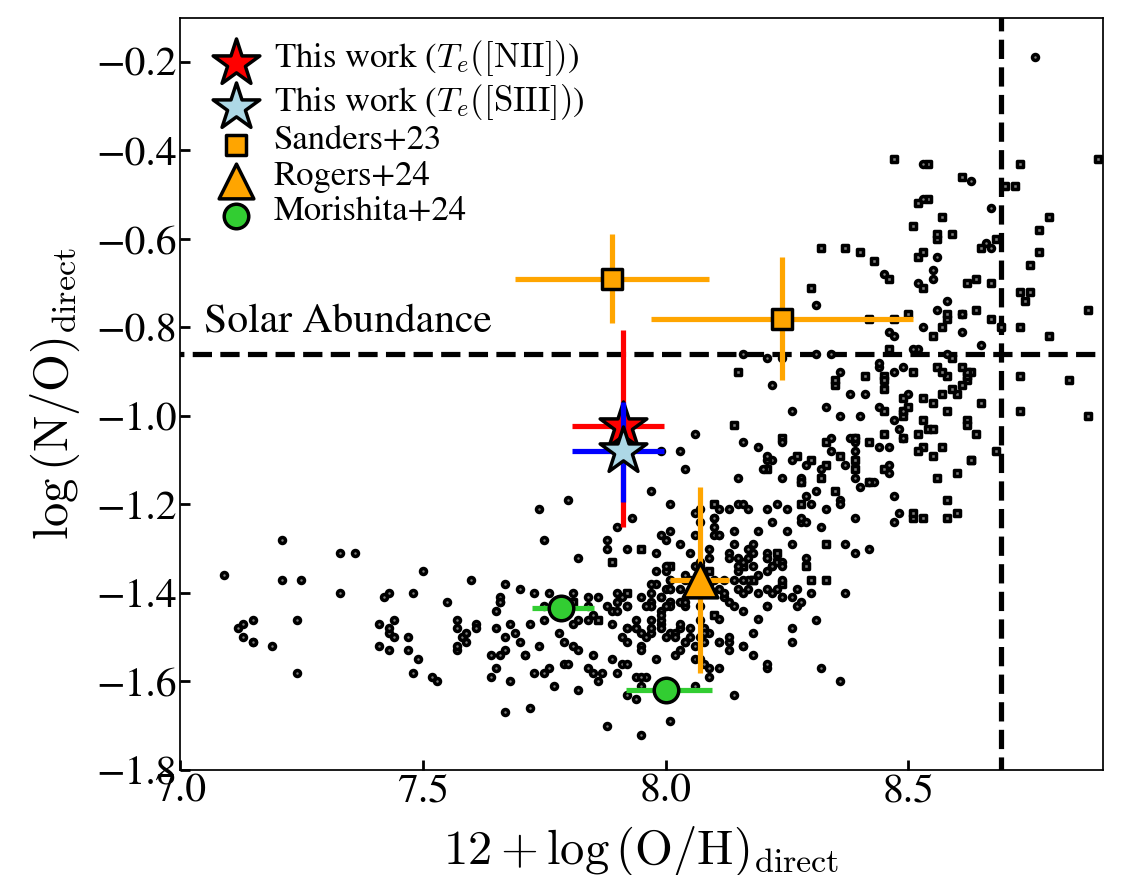}
	\caption{
 \id\ on the (N/O)$_{\rm direct}$-(O/H)$_{\rm direct}$ abundance plane. Also shown are the measurements of \hii\ regions in local galaxies from the CHAOS survey \citep[][black squares]{berg20} and from \citet[][black circles]{pilyugin12}, as well as galaxies at $z\sim2$--3 from recent observations with JWST by \citet[][orange squares]{sanders23} and \citet[][triangle]{rogers23}, and $z=3.4$ and $4.3$ from \citet[][green circles]{morishita24}.
}
\label{fig:abun}
\end{figure}

\subsection{Enhanced Nitrogen-to-Oxygen Abundance}\label{sec:NO}

As shown in Fig.~\ref{fig:abun}, \id\ has a mildly enhanced N/O-abundance compared to local galaxies of the similar O/H abundance \citep[e.g.,][]{vanzee98,pilyugin12,berg20}. In a typical star forming galaxy, nitrogen enrichment predominantly originates from intermediate-mass stars. This occurs on longer timescales than that of $\alpha$ elements, which primarily come from core-collapse supernovae. As such, the observed high N/O abundance is inconsistent with the standard model. 

In the literature, an enhanced N/O abundance has been seen in several individual galaxies \citep[e.g.,][]{teplitz00,sanders23} and as a systematic offset of galaxy populations on the BTP diagram \citep[e.g.,][]{steidel14,masters14}. However, its origin often remains inconclusive, as it can be attributed to multiple scenarios, including ($i$) inflow of pristine gas, lowering O/H abundance but not N/O, ($ii$) the presence of substantial numbers of rapidly rotating massive stars i.e. WR stars \citep[e.g.,][]{vincenzo16,roy21}. In our case, we have detected a broad component in the \heii\,4686 line (Sec.~\ref{sec:heii}). As such, the observed N/O enhancement in \id\ is most likely attributed to the presence of WR stars. 

\id\ has stellar mass of $\logm\sim9.6$. It is unlikely that this amount of stellar mass was formed through a single, short-term burst event --- assuming a constant star formation, at the rate of the inferred SFR (Table~\ref{tab:phys}), it would take $>100$\,Myr to reach the observed mass, which is much longer than the typical time scale of the WR phase for massive stars i.e. $\simlt 1$\,Myr. As such, it is more plausible to assume that \id\ had a star forming phase in the past. In fact, the SFH derived with {\tt gsf} indicates that the $\sim90 \%$ of the observed mass had been formed $\simgt100$\,Myr prior to the observed redshift, or at $z>3$. 

It is fair to note that not all high-$z$ galaxies have an enhanced N/O abundance. In Fig.~\ref{fig:abun}, we also show recent measurements of (N/O)$_{\rm direct}$ with JWST, at $z=2.96$ \citep[][]{rogers23} and $z=3.35$ and $4.25$ \citep{morishita24}. Their N/O abundances are similar to those found in the local galaxies for the corresponding O/H value. As \citet{shapley15} noted, the WR scenario may be unlikely for general galaxy populations, due to the timing requirement to capture the very young stage of the galaxy. Investigating the fraction of WR galaxies with a larger sample will help getting such a timescale estimate.

\begin{figure*}
\centering
    \includegraphics[width=0.46\textwidth]{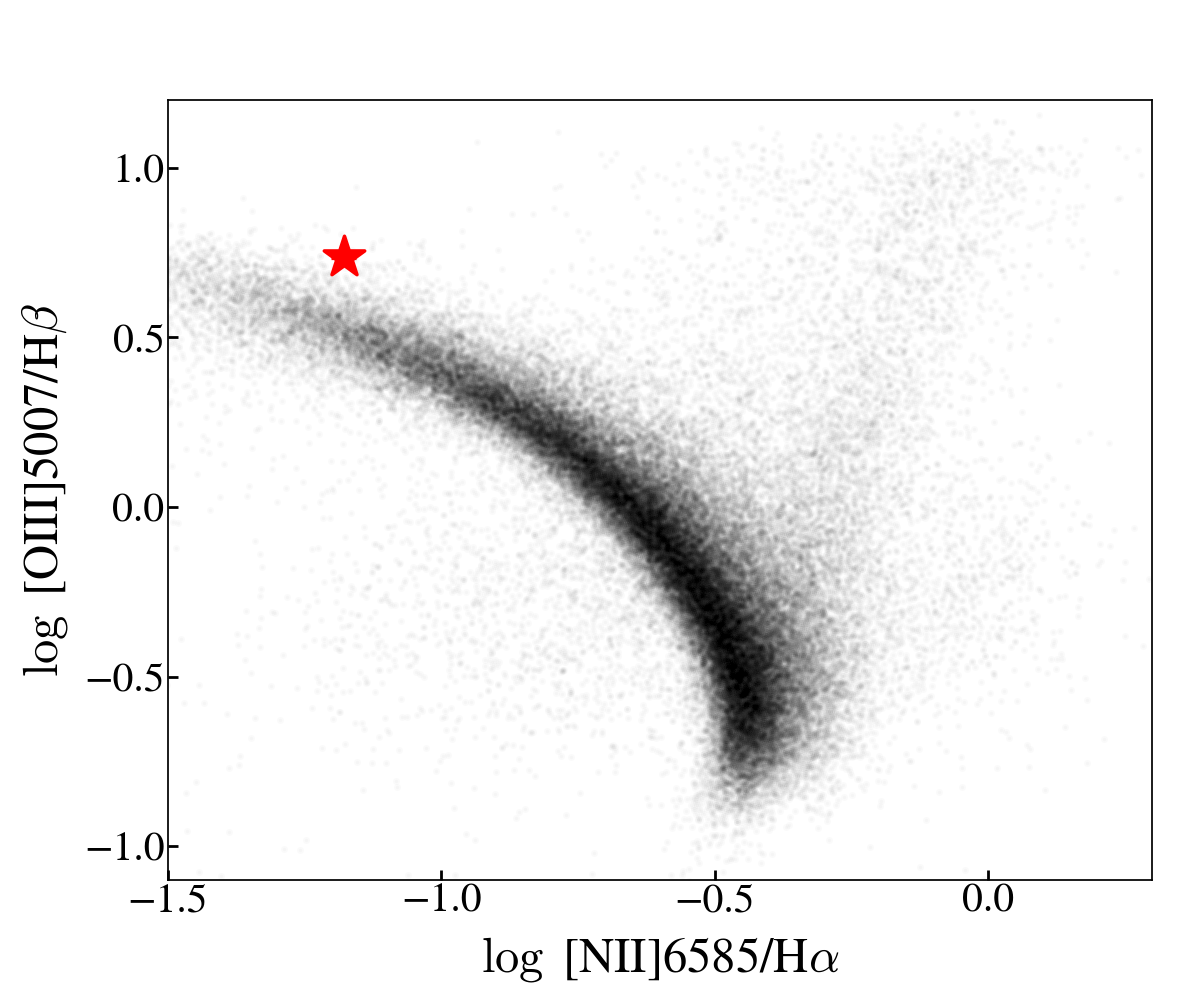}
    \includegraphics[width=0.46\textwidth]{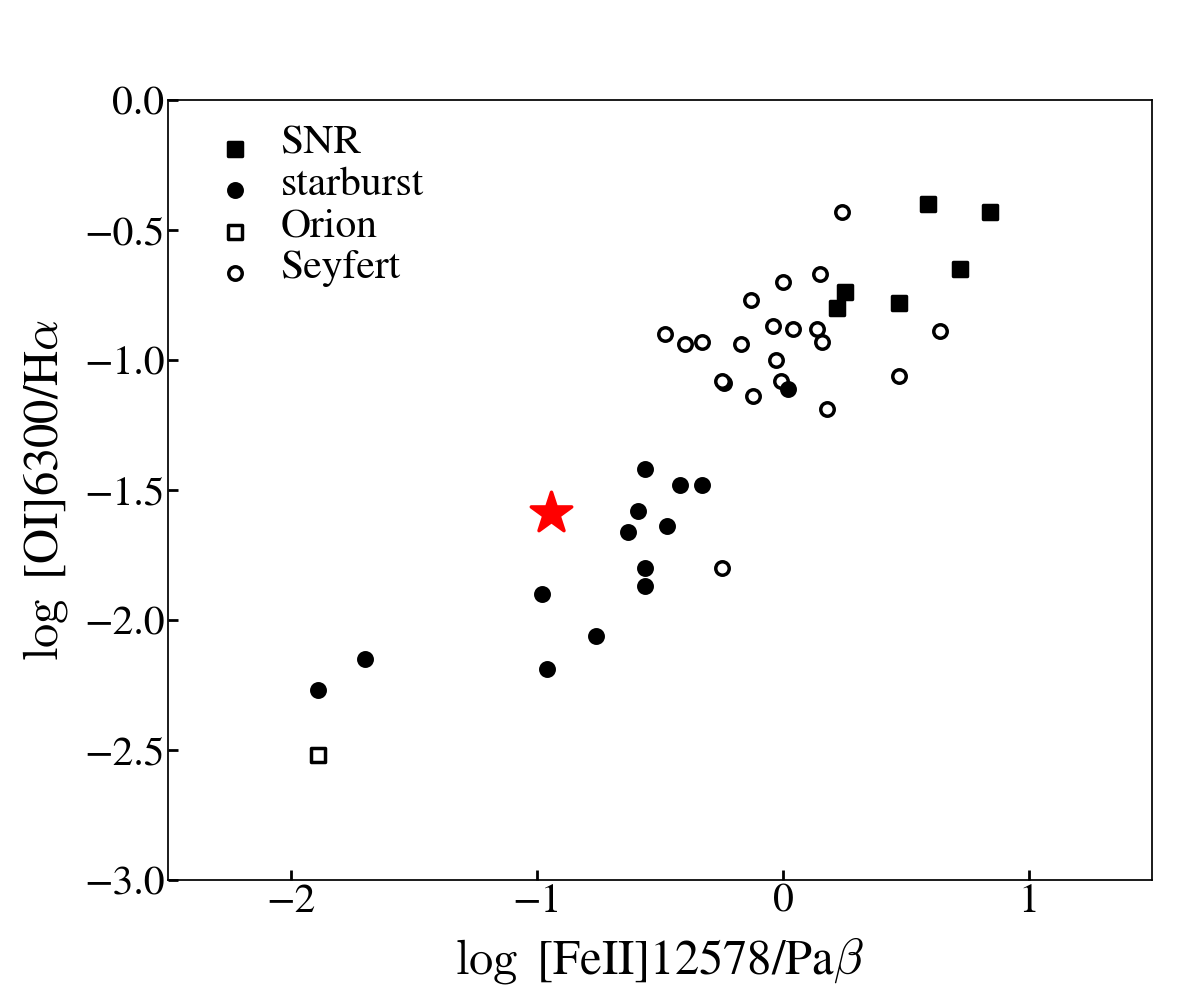}
	\caption{
 (Left): \id\ on the BPT diagram. The location on the diagram indicates that \id\ is not classified as AGN but exhibits hard ionization. Local galaxy measurements, taken from JHU-MPA catalog, are shown (black dots).
 (Right): \id\ on the \feii/\pb--\oi/\ha\ diagram, along with local supernovae remnant (filled squares), Seyfert (open circles), starburst (filled circles), and \hii\ region (Orion; open square) samples taken from \citet{mouri00}. The location of \id\ on this diagram indicates that shock excitation is not required to explain the observed \feii/\pb\ ratio.
 }
\label{fig:Fe2Pab}
\end{figure*}

\subsection{On the absence of AGN}\label{sec:agn}
\id\ is characterized by compact morphology. In addition, we detect the \feii\,1.257 line, which is often seen in AGN or shock-ionized regions around O/B-type stars \citep[e.g.,][]{mouri00}. The presence of AGN could significantly affect the inference of ISM properties, as our analysis implicitly assumes that line emission is primarily driven by photoionization {by stellar radiation}. At the first glance of the spectrum, the scenario is unlikely, as we do not detect any high ionization lines {nor see any broad lines}. Here we further investigate this by using multiple emission line diagnostics. 

The first is the BPT diagram \citep{baldwin81,kewley06}, shown in Fig.~\ref{fig:Fe2Pab}. \id\ is located on the left-top locus of the diagram, near the sequence of local star-forming galaxies. Its location (i.e. high ${\rm O3N2}\equiv\log {\rm (\oiii/H\beta)}-\log {\rm (\nii/H\alpha)}$) indicates a hard radiation field and a low O/H abundance, commonly seen at similar redshifts \citep[e.g.,][]{strom17}. A consistent trend is observed in other strong line ratio diagnostics, such as the \oi-BPT (\oi/\ha--\oiii/\hb) and the \sii-BPT (\sii/\ha--\oiii/\hb). 

We note that on the BPT diagram \id\ is shifted toward the high \nii/\ha\ ($\sim +0.3$\,dex) compared to the local star-forming galaxies of similar \oiii/\hb\ values. We attribute this to the enhanced N/O abundance as discussed in Sec.~\ref{sec:NO}. 

In the right panel of Fig.~\ref{fig:Fe2Pab}, we show the \feii/\pb--\oi/\ha\ diagram, which is used to diagnose the presence of shock excitation, as well as to discriminate Seyfert, starburst, and \hii\ region \citep[e.g.,][]{mouri00}. \id\ is located in the middle, consistent with local starburst galaxies. Shock dominated regions seen in supernovae remnants, exhibit a high \feii/\pb\ ratio, $>1$, a factor of $\sim10$ higher than the observed value in \id. \citet{izotov16} also investigated the ratio in local 18 low-metallicity blue compact dwarf galaxies and concluded that shock excitation is not required to explain their observed ratio, $\feii/{\rm H\beta} \simlt 0.05$ (corresponding to $\feii/{\rm P\beta} \simlt 0.3$). Recently, \citet{brinchmenn23} found the same ratio exceeding $>0.5$ in three galaxies at $z=1.2$--2.5, observed with JWST. While their values are consistent with the presence of shock-dominated regions, they noted that further diagnostics would be required.

\begin{figure*}
\centering
    \includegraphics[width=0.85\textwidth]{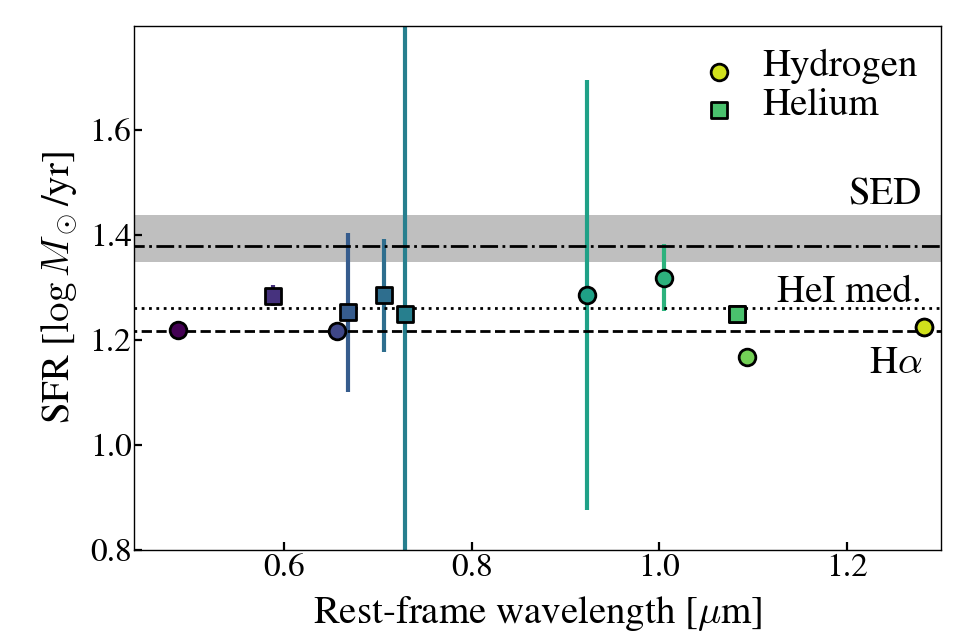}
	\caption{
 Comparison of SFR measured by using various Hydrogen (circles) and \hei\ recombination lines (squares). The mean (equally weighted) \hei\ SFR is shown (dotted line) for comparison with \ha\ SFR (dashed line) and SED-based SFR (dash-dotted line). The excellent agreement indicates that \hei\ lines may be used as an independent SFR indicator.
    }
\label{fig:sfr}
\end{figure*}

\subsection{He~I lines as an independent probe of ISM}\label{sec:he}
We detect multiple \hei\ lines across the observed wavelength range. \hei\ lines are underused in standard ISM analyses, due to their relatively faint nature compared to Hydrogen recombination lines, as well as their complicated aspects coming from collisional excitation and self-absorption. However, the lines could be useful, because ($i$)~the inference via \hei\ is not affected by metallicity, ($ii$)~\hei\ has a higher ionization potential ($24.6$\,eV) than \hi. As such, they can serve as another independent probe of multi-phase ISM, especially in regions dominated by young stars. Here, we proceed our analysis step-by-step. 

We first estimate the electron temperature of \hei. The line ratio $I(7281)/I(6678)$ is known as an excellent temperature indicator, with a small dependence on electron density or optical depth \citep[e.g.,][]{zhang05}. By using the analytic formulae provided by \citet{benjamin99} for the emissivities and $n_e=n_e(\sii)$ (Sec.~\ref{sec:ana}), we find $T_e=1.7_{-0.8}^{+1.3}\times10^4$\,K. Using $I(7281)/I(5876)$, another temperature indicator, we find a similar value, $T_e=2.0_{-0.8}^{+1.0}\times10^4$\,K. The inferred values are higher than those in Sec.~\ref{sec:metal}, though the uncertainty is relatively large, due to the lower S/Ns of \hei\ lines. 

We then estimate optical depth, $\tau_{3889}$, by using the line ratio $I(7065)/I(7281)$. This is due to self-absorption of the meta-stable triplet, which primarily affects the ortho lines (3889, 4471, 5876, 7065, 10830\,{\rm \AA}). While \hei\,7065 is significantly sensitive to such effect, \hei\,7281 is almost insensitive \citep[$\Delta I(7281) <8\,\%$ up to $\tau\sim100$;][]{benjamin02}. Using the electron temperature derived above and the same electron density, we obtain {the optical depth at rest-frame 3889\,\AA\ $\tau_{3889}=2.8\pm0.8$.} We then correct \hei\ line fluxes for the optical depth.

Lastly, we derive nebular extinction using the $\tau$-corrected \hei\ fluxes, in the same manner as in Sec.~\ref{sec:ext}. Even though the fluxes had been corrected for extinction beforehand using Hydrogen recombination lines in Sec.~\ref{sec:ext}, we here find additional extinction of $\Delta\sim0.1$\,mag. 
indicating that \hei\ may be probing a slightly more attenuated region. 

The results above can be explained by the fact that \hei\ has a higher ionization potential, and that the lines predominantly originate in the central region of the galaxy, whereas the \hii\ region is relatively more extended \citep[e.g.,][]{osterbrock89}, to the extent where \oi\ is more dominant than \oii\ and \oiii, as revealed in our analysis in Sec.~\ref{sec:neut}. Since \id\ exhibits compact morphology, more enhanced dust is expected in its central region, where intense massive star formation is likely ongoing, as evidenced by the presence of \heii\ line complex (Sec.~\ref{sec:heii}).

With the additional ISM diagnostics obtained with \hei, we now evaluate its fidelity by comparing star formation rates derived by each of the \hei\ lines with those derived by Hydrogen recombination lines. We note that the conclusion in Sec~\ref{sec:agn} assures that the measured fluxes of Hydrogen and Helium lines are mostly dominated by recombination i.e. star formation. 

We use the \hei\ Case B coefficients used in the analysis above, and the intensity ratio between \hei~5867 and \ha\ assuming the Helium abundance of $Y=0.252$ for the corresponding O/H \citep[e.g.,][]{izotov14}, i.e. $n_{\rm He^+}/n_{\rm H^+}=0.0852$, or equivalently $I_{5867}/I_{\rm H\alpha}=0.040$. This scaling is roughly consistent with observations of local galaxies. In the Sloan DR7 JHU-MPA catalog, we find the average scaling of $I_{5876} / I_{\rm H\alpha} \sim 0.033$. The smaller value found with the JHU-MPA catalog is likely due to the fact that the correction for $\tau_{3889}$ was not applied to their flux measurements.

\begin{figure*}
\centering
    \includegraphics[width=0.45\textwidth]{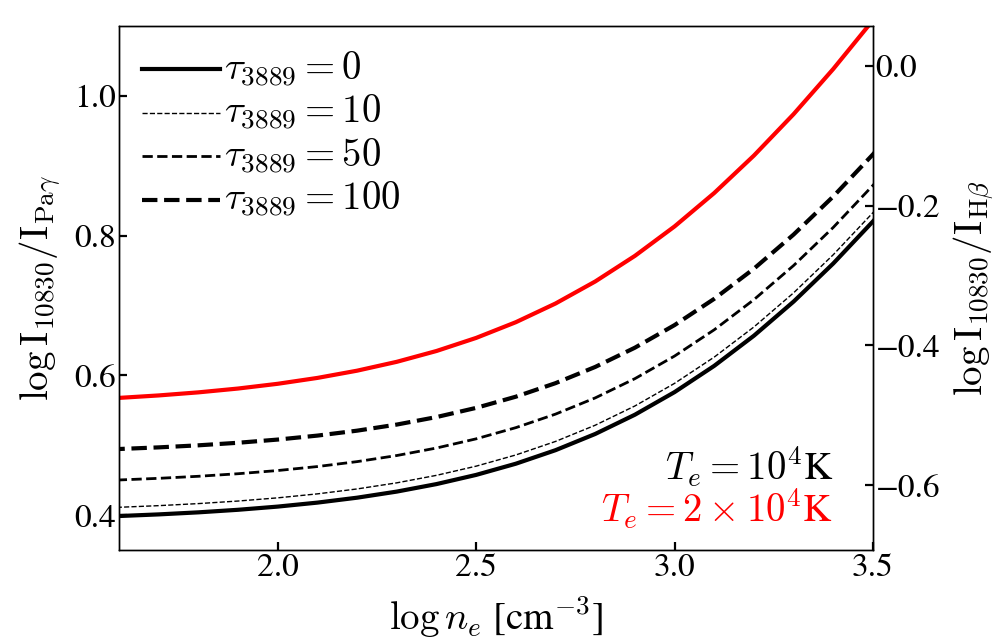}
    \includegraphics[width=0.45\textwidth]{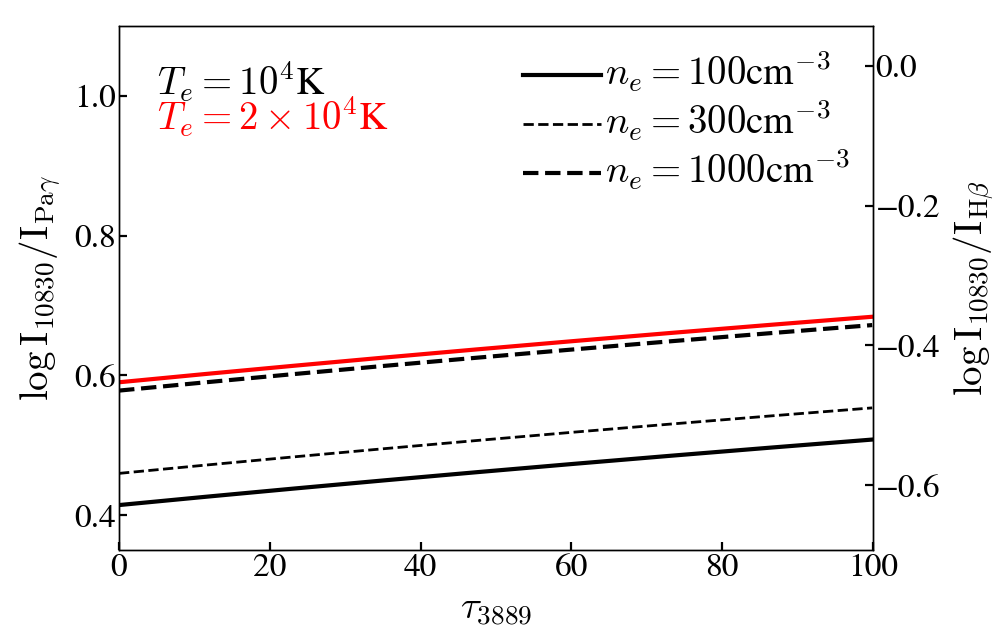}
	\caption{
(Left): Theoretical predictions for the line ratio \hei\,10830/\pg\ for $T_e=10^4$\,K and various \hei\,3889 optical depth $\tau_{3889}$, as a function of electron density ($n_e$; black lines). The prediction for $T_e=2\times10^4$\,K and $\tau_{3889}=0$ is also shown.
(Right): Same ratio is shown but as a function of \hei optical depth $\tau_{3889}$ for various electron density. The prediction for $T_e=2\times10^4$\,K and $n_e=100$\,cm$^{-3}$ is also shown.
    }
\label{fig:HeIrat}
\end{figure*}

In Fig.~\ref{fig:sfr}, we show the SFRs derived with various lines and one from the SED analysis in Sec.~\ref{sec:sed}.
Across the wavelength range, we see an excellent agreement, with SFRs from \hei\ being higher but only $\simlt 0.1$\,dex, and the difference is still smaller than the one derived from SED fitting. 
The agreement seen in Fig.~\ref{fig:sfr} is encouraging, and demonstrates a potential use case of \hei\ lines as an alternative SFR indicator. {Especially, at the rest-frame wavelength of $\sim1.1\,\mu$m, the \hei\,10830 is among the brightest, and even much brighter than the adjacent Pa$\gamma$. (The observed ratio for \id\ is $I_{10830}/I_{\rm Pa\gamma}=4.8$.) Fig.~\ref{fig:HeIrat} shows the theoretical predictions for the line intensity normalized by \pg\ and \hb.}
{In addition, the line is observable with JWST NIRSpec over a wide range of redshift, $0\simlt z\simlt 4$. With these aspects, we see that future studies may benefit from the line. For convenience, we provide the following formula for the \hei\,10830--SFR conversion:}
\begin{equation}
    {\rm SFR} / M_\odot {\rm yr^{-1}} = \left( \frac{C}{0.0852\,Y} \right) \left( \frac{L_{\hei, \lambda 10830}}{2 \times10^{41}\,{\rm erg\,s^{-1}}} \right),
\end{equation}
{for the Chabrier IMF}, where the conversion coefficient $C$ includes the effect of $n_e$, $T_e$, and $\tau_{3889}$. $Y$ is the Helium mass fraction, which in this study we derived with $Y = 0.25 + 26.13 \times10^{\log {\rm (O/H)}}$ \citep{izotov14}. We have tabulated the conversion coefficients for various parameters in Table~\ref{tab:hei_sfr}.

However, care must be taken. Our analysis involves multiple \hei\ lines, as well as $n_e$ inferred with \sii. All these have enabled us to robustly obtain $T_e$, $\tau$, and nebular extinction of \hei. The \hei\,10830 line is meta-stable and subject to self-absorption. For example, the line ratio $I_{10830} / I_{5876}$ can significantly vary as a function of electron density (Fig.~\ref{fig:HeIrat}). Similarly, it has a mild temperature dependence, and the difference can be $\sim40\,\%$ between 10000\,K and 20000\,K.



\section{Summary}\label{sec:sum}

We reported JWST/NIRSpec observations of a star-forming galaxy at $z=2.76$, \id. Our findings are summarized as follows:
\begin{itemize}
    \item We detected three auroral lines, \siii\,6312, \oii\,{7320,7331}, and \nii\,5755.  We found that the electron temperatures derived using those lines are consistent.

    \item \id\, exhibits a broad line complex at rest-frame wavelength of \heii\,4686. Along with the observed high-N/O abundance, we concluded that \id\ is a young galaxy that hosts a significant number of Wolf-Rayet stars.

    \item We diagnosed the presence of AGN by using strong emission lines and concluded that the observed features of \id\ is primarily driven by photoionization, validating our ISM analysis.

    \item We performed another ISM analysis using a series of \hei\ lines, and found that \hei\ is likely probing a smaller scale region than \hi. The inferred properties are broadly consistent with the results above. 
\end{itemize}

With the exceptional sensitivity of JWST, we conducted an in-depth analysis of emission lines. Although our study focused on a single galaxy, ongoing JWST observing programs like CECILIA \citep[][]{strom23} and AURORA \citep{shapley21,shapley24} will soon allow for the precise examination of ISM properties across a large number of galaxies at similar redshifts. Further exploration of galaxies exhibiting evidence of Wolf-Rayet stars, as discovered in \id, would be of great interest. Determining the prevalence of these features across a large number of galaxies will provide critical insights into the timescale of massive star formation and the initial mass function.

In addition, we have demonstrated that \hei\ lines can serve as an independent probe of galaxy ISM. Particularly noteworthy is the strength of the \hei\,10830 line compared to adjacent \pg. This line can be as strong as \hb\ under moderate attenuation, $A_V\simgt1$\,mag. Since this line falls in the wavelength range of the sensitive spectrographs onboard JWST up to $z\sim4$, we anticipate it becoming another reliable calibrator for star formation rate.


\input{table_phys.tex}

\input{helium_sfr_coeff.tex}


\section*{Acknowledgements}
TM would like to thank Lee Armus and Ranga-Ram Chary for useful discussion. 
Some/all of the data presented in this paper were obtained from the Mikulski Archive for Space Telescopes (MAST) at the Space Telescope Science Institute. The specific observations analyzed can be accessed via \dataset[10.17909/q8cd-2q22]{https://doi.org/10.17909/q8cd-2q22}. We acknowledge support for this work under NASA grant 80NSSC22K1294. SS has received funding from the European Union’s Horizon 2022 research and innovation programme under the Marie Skłodowska-Curie grant agreement No 101105167 - FASTIDIoUS.

{
{\it Software:} 
Astropy \citep{astropy13,astropy18,astropy22}, bbpn \citep{bbpn}, EAzY \citep{brammer08}, EMCEE \citep{foreman13}, gsf \citep{morishita19}, numpy \citep{numpy}, python-fsps \citep{foreman14}.
}


\bibliography{ms}{}
\bibliographystyle{aasjournal}



\end{document}

%% file: table_linefluxes.tex
\begin{deluxetable*}{ccccc}
\tabletypesize{\footnotesize}
\tablewidth{-2pt}
\tablecaption{
Emission line measurements of \id.
}
\tablehead{
\colhead{Line} & \colhead{IP} & \colhead{Flux} & \colhead{$\sigma$} & \colhead{EW$_0$}\\
\colhead{} & \colhead{eV} & \colhead{$10^{-19}$\,erg/s/cm$^2$} & \colhead{km/s} & \colhead{$\mathrm{\AA}$}
}
\startdata
${\rm HeII}_{4686}$ & 54.4 & $39.93 \pm 2.25$ & $1347.4 \pm 310.5$ & $37.8 \pm 6.0$\\
${\rm [ArIV]}_{4711}$ & 40.74 & $<1.16$ & -- & --\\
${\rm HeI}_{4714}$ & 24.6 & $3.11 \pm 0.96$ & $223.3 \pm 126.9$ & $2.4 \pm 1.6$\\
${\rm [ArIV]}_{4740}$ & 40.74 & $3.74 \pm 1.02$ & $282.0 \pm 83.2$ & $2.3 \pm 1.5$\\
${\rm H\beta}_{4861}$ & 3.4 & $174.50 \pm 1.83$ & $<164.3$ & $130.3 \pm 1.0$\\
${\rm HeI}_{4922}$ & 24.6 & $<2.07$ & -- & --\\
${\rm [OIII]}_{4959+5007}$ & 35.1 & $1271.14 \pm 9.80$ & $<161.2$ & $228.0 \pm 0.7$\\
${\rm [OI]}_{5577}$ & 0 & $<1.46$ & -- & --\\
${\rm [NII]}_{5755}$ & 14.5 & $1.12 \pm 0.47$ & $<142.8$ & $1.1 \pm 0.3$\\
${\rm HeI}_{5876}$ & 24.6 & $21.34 \pm 0.68$ & $142.9 \pm 4.4$ & $23.8 \pm 0.7$\\
${\rm [OI]}_{6300}$ & 0 & $10.15 \pm 0.32$ & $151.9 \pm 6.5$ & $14.0 \pm 0.5$\\
${\rm [SIII]}_{6312}$ & 19.1 & $3.04 \pm 0.38$ & $154.8 \pm 27.6$ & $4.1 \pm 0.6$\\
${\rm [OI]}_{6364}$ & 0 & $2.89 \pm 0.09$ & $151.9 \pm 6.5$ & $4.0 \pm 0.1$\\
${\rm SiII}_{6373}$ & 8.2 & $1.44 \pm 0.41$ & $204.0 \pm 42.5$ & $2.1 \pm 0.6$\\
${\rm [NII]}_{6548+6583}$ & 14.5 & $52.52 \pm 2.66$ & $150.7 \pm 2.2$ & $29.3 \pm 0.4$\\
${\rm H\alpha}_{6563}$ & 3.4 & $499.15 \pm 3.55$ & $142.7 \pm 0.4$ & $750.2 \pm 2.3$\\
${\rm HeI}_{6678}$ & 24.6 & $5.33 \pm 0.33$ & $128.8 \pm 11.2$ & $7.5 \pm 0.5$\\
${\rm [SII]}_{6716}$ & 10.4 & $24.90 \pm 0.42$ & $137.0 \pm 2.8$ & $36.0 \pm 0.7$\\
${\rm [SII]}_{6731}$ & 10.4 & $22.23 \pm 0.41$ & $133.7 \pm 3.0$ & $32.4 \pm 0.6$\\
${\rm OI}_{7002}$ & 0 & $<0.86$ & -- & --\\
${\rm HeI}_{7065}$ & 24.6 & $8.34 \pm 0.41$ & $126.5 \pm 6.6$ & $14.5 \pm 0.7$\\
${\rm [ArIII]}_{7136}$ & 27.6 & $12.24 \pm 0.48$ & $136.9 \pm 5.2$ & $22.5 \pm 0.8$\\
${\rm HeI}_{7281}$ & 24.6 & $1.52 \pm 0.46$ & $172.3 \pm 36.0$ & $2.8 \pm 0.8$\\
${\rm [OII]}_{7320}$ & 13.6 & $4.91 \pm 0.41$ & $111.2 \pm 12.6$ & $9.5 \pm 0.9$\\
${\rm [OII]}_{7331}$ & 13.6 & $5.04 \pm 0.46$ & $149.6 \pm 18.9$ & $9.4 \pm 1.1$\\
${\rm [NiII]}_{7378}$ & 7.6 & $<0.80$ & -- & --\\
${\rm [NiII]}_{7411}$ & 7.6 & $<0.88$ & -- & --\\
${\rm [ArIII]}_{7751}$ & 27.6 & $1.72 \pm 0.43$ & $<102.6$ & $3.9 \pm 0.7$\\
${\rm OI}_{7774}$ & 0 & $<1.05$ & -- & --\\
${\rm HeI}_{7816}$ & 24.6 & $<1.14$ & -- & --\\
${\rm HeII}_{8237}$ & 54.4 & $<0.64$ & -- & --\\
${\rm [SIII]}_{9069}$ & 19.1 & -- & -- & --\\
${\rm Pa9}_{9229}$ & 1.5 & $3.76 \pm 0.44$ & $175.9 \pm 19.0$ & $12.5 \pm 1.0$\\
${\rm [SIII]}_{9531}$ & 19.1 & $55.88 \pm 0.73$ & $158.0 \pm 1.5$ & $203.5 \pm 1.7$\\
${\rm Pa\delta}_{10049}$ & 1.5 & $8.08 \pm 0.29$ & $<133.5$ & $29.7 \pm 1.0$\\
${\rm HeI}_{10830}$ & 24.6 & $66.32 \pm 0.53$ & $155.6 \pm 1.1$ & $263.6 \pm 1.8$\\
${\rm Pa\gamma}_{10938}$ & 1.5 & $14.04 \pm 0.31$ & $123.7 \pm 2.5$ & $60.0 \pm 1.1$\\
${\rm OI}_{11287}$ & 0 & $1.00 \pm 0.23$ & $162.7 \pm 18.0$ & $5.0 \pm 1.2$\\
${\rm [FeII]}_{12567}$ & 7.9 & $3.33 \pm 0.22$ & $142.9 \pm 11.9$ & $18.0 \pm 1.7$\\
${\rm HeI}_{12790}$ & 24.6 & $1.90 \pm 0.39$ & $133.6 \pm 16.2$ & $8.8 \pm 1.3$\\
${\rm Pa\beta}_{12818}$ & 1.5 & $29.20 \pm 0.52$ & $121.1 \pm 1.8$ & $135.4 \pm 1.8$\\
\enddata
\tablecomments{
Fluxes are in units of $10^{-19}$\,erg/s/cm$^2$. Flux errors are $1\,\sigma$. $2\,\sigma$ upper limits are quoted for those undetected (S/N\,$<2$). Attenuation has been corrected using $A_V=0.43$.
EW$_0$: Rest-frame equivalent width.
}\label{tab:lines}
\end{deluxetable*}

%% file: table_phys.tex
\begin{deluxetable*}{ccc}
\tabletypesize{\footnotesize}
\tablecolumns{3}
\tablewidth{0pt}
\tablecaption{
Physical properties of \id}
\tablehead{
\colhead{Property} & \colhead{Unit} & \colhead{Measurement}\\
\colhead{} & \colhead{} & \colhead{}
}
\startdata
\cutinhead{Host properties}
R.A. & degree & $1.773851e+02$\\
Dec. & degree & $2.237533e+01$\\
$M_*$ & $\log M_\odot$ & $9.6_{-0.1}^{+0.2}$\\
$A_{V, \mathrm{SED}}$ & mag & $0.3_{-0.1}^{+0.1}$\\
$\beta_{\mathrm{UV}}$ &  & $-1.7_{-0.1}^{+0.1}$\\
$R_e$ & kpc & $0.7_{-0.1}^{+0.1}$\\
\cutinhead{ISM properties}
$n_{e}(\mathrm{[SII]})$ & $10^{2}$cm$^{-3}$ & $3.63_{-0.64}^{+0.76}$\\
$T_{e}(\mathrm{[SIII]})$ & $10^{4}$\,K & $1.47_{-0.14}^{+0.17}$\\
$T_{e}(\mathrm{[NII]})$ & $10^{4}$\,K & $1.37_{-0.25}^{+0.38}$\\
$T_{e}(\mathrm{[OII]})_\mathrm{[OIII]}$ & $10^{4}$\,K & $1.39_{-0.07}^{+0.07}$\\
$T_{e}(\mathrm{[OIII]})_\mathrm{[NII]}$ & $10^{4}$\,K & $1.47_{-0.33}^{+0.45}$\\
$T_{e}(\mathrm{[OIII]})_\mathrm{[SIII]}$ & $10^{4}$\,K & $1.50_{-0.13}^{+0.19}$\\
$T_{e}(\mathrm{[OI]})$ & $10^{4}$\,K & $<1.99$\\
$12 + \mathrm{\log (O^+/H^+)}$ &  & $7.33_{-0.04}^{+0.04}$\\
$12 + \mathrm{\log (O^{2+}/H^+)}$ &  & $7.78_{-0.13}^{+0.10}$\\
$12 + \mathrm{\log (O/H)}$ &  & $7.91_{-0.10}^{+0.08}$\\
$\mathrm{\log (S^+/H^+)}$ &  & $5.50_{-0.01}^{+0.01}$\\
$\mathrm{\log (S^{2+}/H^+)}$ &  & $5.98_{-0.07}^{+0.07}$\\
$\mathrm{\log (S/O)}$ &  & $-1.75_{-0.05}^{+0.06}$\\
$\mathrm{\log (N^+/H^+)}$ &  & $6.31_{-0.23}^{+0.22}$\\
$\mathrm{\log (N/O)_\mathrm{[SIII]}}$ &  & $-1.08_{-0.11}^{+0.11}$\\
$\mathrm{\log (N/O)_\mathrm{[NII]}}$ &  & $-1.02_{-0.23}^{+0.22}$\\
$\mathrm{\log (Ar^{2+}/H^+)}$ &  & $5.47_{-0.02}^{+0.02}$\\
$\mathrm{\log (Ar/O)}$ &  & $-2.40_{-0.02}^{+0.02}$\\
$n(\mathrm{H^+})/n(\mathrm{H^0})$ &  & $242$\\
\cutinhead{ISM properties (He~I)}
$T_{e}(\mathrm{HeI,6678})$ & $10^{4}$\,K & $2.01_{-0.82}^{+0.98}$\\
$T_{e}(\mathrm{HeI,5876})$ & $10^{4}$\,K & $1.74_{-0.78}^{+1.25}$\\
$\tau_{3889}$ &  & $2.80_{-0.70}^{+0.80}$\\
$A_{V, \mathrm{HeI, SMC}}$ & mag & $0.5$\\
\cutinhead{Starformation rates}
$\mathrm{SFR_{SED}}$ & $\log M_\odot {\rm yr^{-1}}$ & $1.33_{-0.03}^{+0.03}$\\
$\mathrm{SFR_{Hb4861}}$ & $\log M_\odot {\rm yr^{-1}}$ & $1.06 \pm {0.01}$\\
$\mathrm{SFR_{Ha6563}}$ & $\log M_\odot {\rm yr^{-1}}$ & $1.06 \pm {0.01}$\\
$\mathrm{SFR_{Pa99229}}$ & $\log M_\odot {\rm yr^{-1}}$ & $1.13 \pm {0.41}$\\
$\mathrm{SFR_{Pad10049}}$ & $\log M_\odot {\rm yr^{-1}}$ & $1.16 \pm {0.06}$\\
$\mathrm{SFR_{Pag10938}}$ & $\log M_\odot {\rm yr^{-1}}$ & $1.01 \pm {0.02}$\\
$\mathrm{SFR_{Pab12818}}$ & $\log M_\odot {\rm yr^{-1}}$ & $1.07 \pm {0.01}$\\
$\mathrm{SFR_{HeI5876}}$ & $\log M_\odot {\rm yr^{-1}}$ & $1.13 \pm {0.02}$\\
$\mathrm{SFR_{HeI6678}}$ & $\log M_\odot {\rm yr^{-1}}$ & $1.10 \pm {0.15}$\\
$\mathrm{SFR_{HeI7065}}$ & $\log M_\odot {\rm yr^{-1}}$ & $1.13 \pm {0.11}$\\
$\mathrm{SFR_{HeI7281}}$ & $\log M_\odot {\rm yr^{-1}}$ & $1.09 \pm {2.47}$\\
$\mathrm{SFR_{HeI10830}}$ & $\log M_\odot {\rm yr^{-1}}$ & $1.09 \pm {0.01}$\\
\enddata
\tablecomments{
Measurements are corrected for magnification using $\mu=1.34$ (Sec.~\ref{sec:data}). Quoted errors are $1\,\sigma$. 
}\label{tab:phys}
\end{deluxetable*}

%% file: helium_sfr_coeff.tex
\begin{deluxetable*}{cccccc}
\tabletypesize{\footnotesize}
\tablewidth{-2pt}
\tablecaption{
The \hei\,10830--SFR conversion coefficients.
}
\tablehead{
\colhead{$T_e$} & \colhead{$\tau_{3889}$} & \multicolumn{4}{c}{$C$}\\
\cline{3-6}
\colhead{$K$} & \colhead{} & \colhead{$n_e=100 {\rm cm^{-3}}$} & \colhead{$n_e=300 {\rm cm^{-3}}$} & \colhead{$n_e=1000 {\rm cm^{-3}}$} & \colhead{$n_e=3000 {\rm cm^{-3}}$}
}
\startdata
5000 & 0 & 1.2437 & 1.1949 & 1.0507 & 0.7818\\
5000 & 10 & 1.2363 & 1.1878 & 1.0445 & 0.7772\\
5000 & 100 & 1.0755 & 1.0333 & 0.9086 & 0.6761\\
10000 & 0 & 1.0384 & 0.9426 & 0.7125 & 0.4198\\
10000 & 10 & 1.0323 & 0.9370 & 0.7083 & 0.4173\\
10000 & 100 & 0.8980 & 0.8151 & 0.6161 & 0.3630\\
15000 & 0 & 0.8779 & 0.7751 & 0.5497 & 0.3003\\
15000 & 10 & 0.8727 & 0.7705 & 0.5465 & 0.2985\\
15000 & 100 & 0.7592 & 0.6703 & 0.4754 & 0.2597\\
20000 & 0 & 0.7659 & 0.6656 & 0.4564 & 0.2404\\
20000 & 10 & 0.7614 & 0.6617 & 0.4537 & 0.2390\\
20000 & 100 & 0.6624 & 0.5756 & 0.3947 & 0.2079\\
\enddata
\tablecomments{
${\rm SFR} / M_\odot {\rm yr^{-1}} = \left( \frac{C}{0.0852 Y} \right) \left( \frac{L_{\hei,{\lambda10830}}}{2 \times10^{41} {\rm erg s^{-1}}} \right)$. Coefficients for a wider range of parameters can be found at \url{https://github.com/mtakahiro/Helium}.
}\label{tab:hei_sfr}
\end{deluxetable*}